\documentclass[a4paper,11pt]{article}
\pdfoutput=1

\usepackage{CTPpub}
\usepackage[T1]{fontenc}
\usepackage{multirow}
\usepackage{caption}
\usepackage{subcaption}

\title{\boldmath Precision measurements and tau neutrino physics in a future accelerator neutrino experiment}

\author{Jian Tang,}
\author{Sampsa Vihonen}
\author{and Yu Xu}

\affiliation{School of Physics, Sun Yat-sen University, Guangzhou 510275, China}

\emailAdd{tangjian5@mail.sysu.edu.cn}
\emailAdd{sampsa@mail.sysu.edu.cn}
\emailAdd{xuyu27@mail.sysu.edu.cn}

\abstract{We investigate prospects of building a future accelerator-based neutrino oscillation experiment in China, including site selection, beam optimization and tau neutrino physics aspects. {\em CP} violation, non-unitary mixing and non-standard neutrino interactions are discussed. We simulate neutrino beam setups based on muon and beta decay techniques and compare Chinese laboratory sites by their expected sensitivities. A case study on Super Proton-Proton Collider and China JinPing Laboratory is also presented.  {\color{black}It is shown that} the muon-decay-based beam setup can measure the Dirac {\em CP} phase by about {\color{black}14.2$^\circ$ precision at 1$\,\sigma$ CL}, whereas non-unitarity can be probed down to $|\alpha_{i j}| \lesssim$ 0.37 ($i \neq j =$ 1, 2, 3) and non-standard interactions to $|\epsilon^m_{\ell \ell'}| \lesssim$ 0.11 ($\ell \neq \ell' = e$, $\mu$, $\tau$) {\color{black}at 90\% CL}, respectively.}


\begin{document} 
\maketitle
\flushbottom

\section{Introduction}
\label{sec:intro}

The near-future of neutrino oscillation physics will be highlighted by precision measurements on the neutrino oscillation parameters. The missing pieces of the mechanism that governs the oscillations in the three-neutrino picture will be searched in a variety of neutrino oscillation experiments. The next generation of neutrino experiments, including the Jiangmen Underground Neutrino Observatory (JUNO)~\cite{An:2015jdp}, Tokai-to-Hyper-Kamiokande (T2HK)~\cite{Abe:2015zbg} and Deep Underground Neutrino Experiment (DUNE)~\cite{Acciarri:2015uup}, will look for the remaining unknowns in the Pontecorvo-Maki-Nakagawa-Sakata (PMNS) matrix~\cite{Pontecorvo:1957cp,Pontecorvo:1957qd,Maki:1960ut,Maki:1962mu,Pontecorvo:1967fh} and squared differences\footnote{The main objectives are to determine whether neutrino masses follow normal ordering, $m_1 < m_2 < m_3$, or inverted ordering, $m_3 < m_1 < m_2$, and whether the Dirac {\em CP} phase $\delta_{\rm CP}$ is {\em CP}-conserving, $\sin \delta_{\rm CP} \neq$ 0 or {\em CP}-violating, $\sin \delta_{\rm CP} \neq$ 0. Also waiting to be measured is the octancy of $\theta_{23}$, which may be either low, $\theta_{23} <$ 45$^\circ$, or high, $\theta_{23} >$ 45$^\circ$.}. These experiments will furthermore improve the precision on the individual parameters and provide probes to new physics\footnote{The prospects of using future experiments to search new physics have been studied extensively in the literature. Some good examples are Refs.~\cite{Palazzo:2020tye,Du:2021rdg,Bakhti:2020fde,Abi:2020kei,C.:2019dbf,Fukasawa:2016lew,Huitu:2016bmb}.}. It is especially important to conduct measurements on the oscillations involving tau neutrinos in order to improve the precision of unitarity tests on the PMNS matrix~\cite{Hu:2020oba}. Without answers are also the questions of absolute scale of neutrino masses and whether neutrinos are of Dirac or Majorana nature, which are studied in neutrinoless double beta decay searches and cosmological probes as well as in direct neutrino mass experiments such as the Karlsruhe TRItium Neutrino experiment (KATRIN)~\cite{Osipowicz:2001sq}.

China has been a major stage to reactor neutrino physics for many years. The most notable accomplishments were achieved in the Daya Bay reactor neutrino experiment, which contributed to the discovery of the non-zero reactor mixing angle $\theta_{13}$~\cite{Guo:2007ug}. In the next few years, the analysis of the reactor neutrino data will be continued in JUNO, with the main goal set in determining the neutrino mass ordering by at least 3$\,\sigma$ confidence level (CL)~\cite{An:2015jdp} {\color{black}within 6 years of data}. At the same time, the discovery of dark matter is being sought in the China JinPing Laboratory (CJPL)~\cite{Cao:2014jsa} with PandaX~\cite{Aprile:2015uzo} and CDEX~\cite{Kang:2013sjq}. There are also a number of accelerator laboratories being constructed in China, which offer an opportunity to consider an accelerator-based experimental neutrino program. One promising example is the proposal of MuOn-decay MEdium-baseline neutrino facility (MOMENT)~\cite{Cao:2014bea}, which presents a novel concept of producing a high-intensity low-energy neutrino beam using muon decay as their source. The physics case of MOMENT has been established in both the CP-violation searches and in precision measurements of the standard neutrino oscillation parameters as well as in the searches for new physics~\cite{Blennow:2015cmn,Bakhti:2016prn,Tang:2017qen,Tang:2018rer,Tang:2019edw,Tang:2019wsv,Tang:2020xoy}. On the other end of the national research planning in China are the next-generation collider experiments of the Circular Electron-Positron Collider (CEPC) and its high-energy upgrade Super Proton-Proton Collider (SPPC)~\cite{Chou:2015noa,CEPC-SPPCStudyGroup:2015csa,CEPC-SPPCStudyGroup:2015esa}. 

We investigate the prospects of building a future accelerator-based neutrino oscillation experiment in China. In this physics-driven study, we discuss the possibilities of the Chinese landscape to study neutrino oscillations in a long-baseline experiment. The neutrino detector, for example, could be placed in an underground laboratory such as JUNO or CJPL. The neutrino beamline on the other hand could be built in an existing accelerator laboratory such as the China Spallation Neutron Source (CSNS)~\cite{CSNS2006} and China initiative for Accelerator Driven System (CiADS)~\cite{CiADS2012}, or in a planned facility such as the one at the Institute of Modern Physics of the Chinese Academy of Sciences (CAS-IMP)~\cite{Mao:2020rlb} or at the Proton Linear Accelerator Institute in Nanjing University~\cite{Sun:2017}. Another interesting option is to build a neutrino beam facility at the SPPC accelerator complex, where the proton beam required for neutrino production could be diverted from one of the accelerator rings in the injector chain. In the present work, we briefly review the available locations for the accelerator and detector facilities and analyse their location-based physics prospects in establishing the conservation or violation of the CP symmetry, determining the value of the Dirac {\em CP} phase as well as other standard oscillation parameters, and running precision tests on Standard Model by searching for signs of non-standard interactions and non-unitary neutrino mixing. We study the experiment sensitivities for the available baselines by simulating a hypothetical accelerator neutrino experiment with GLoBES~\cite{Huber:2004ka,Huber:2007ji}. In this regard, we consider two neutrino beam techniques, where neutrino production is driven by either muon decay or beta decay. We also present a case study on a specific configuration where an accelerator neutrino facility based on muon-decay-based neutrino production~\cite{Huber:2002mx,Autiero:2003fu,Huber:2006wb,Choubey:2010zz,Choubey:2011zzq} is established at the SPPC injector chain whereas neutrino detector similar to magnetized iron and emulsion chamber technologies is constructed at CPJL site, introducing a baseline length of 1736 km. Noting also the prospects of studying tau neutrinos in long-baseline neutrino experiments~\cite{Kopp:2008ds,Meloni:2019pse,DeGouvea:2019kea,Abi:2020kei}, we place emphasis on $\nu_\tau$ appearance. We call this setup {\em {\bf\em PR}ecisi{\bf\em O}n {\bf\em M}easurements and {\bf\em P}hysics with {\bf\em T}au neutrinos} ({\bf PROMPT}) and investigate its prospects as a future accelerator neutrino experiment.

The article is organized as follows: we define the physics scenarios to be considered in this work in section\,\ref{sec:prob}. In section\,\ref{sec:labsites}, we present an overview on several accelerator and underground laboratories in China. In section\,\ref{sec:conf} we discuss the potential experiment setups that can be established in the considered research infrastructure. We then compare their physics prospects in section\,\ref{sec:pheno}. We summarize our findings in section\,\ref{sec:summ}.

\section{Review of theoretical formalism}
\label{sec:prob}

In this section, we briefly examine the theoretical formalism for neutrino oscillations relevant for long-baseline experiments. We begin with the standard scenario with the Pontecorvo-Maki-Nakagawa-Sakata (PMNS) matrix and Mikheyev-Smirnov-Wolfenstein (MSW) effect in section\,\ref{sec:prob1}, extend the discussion to non-unitary mixing in section\,\ref{sec:prob2} and non-standard interactions in section\,\ref{sec:prob3}.

\subsection{The standard paradigm}
\label{sec:prob1}

In the standard parameterization of neutrino oscillations, the mixing between the three active neutrinos $\nu_e$, $\nu_\mu$ and $\nu_\tau$ is described by the Pontecorvo-Maki-Nakagawa-Sakata (PMNS) matrix~\cite{Pontecorvo:1957cp,Pontecorvo:1957qd,Maki:1960ut,Maki:1962mu,Pontecorvo:1967fh}, which is a unitary 3$\times$3 matrix that decomposes into three parts:
\begin{equation}\label{eq:PMNS}
U_{\rm PMNS} =
\left(
\begin{array}{ccc}
1 & 0 & 0\\
0 & c_{23} & s_{23} \\
0 & -s_{23} & c_{23}
\end{array}\right) 
\left(
\begin{array}{ccc}
c_{13} & 0 & s_{13}\mathrm{e}^{-i\delta_\mathrm{CP}}\\
0 & 1 & 0 \\
s_{13}\mathrm{e}^{i\delta_\mathrm{CP}} & 0 & c_{13}
\end{array}\right) 
\left(
\begin{array}{ccc}
c_{12} & s_{12} & 0\\
-s_{12} & c_{12} & 0 \\
0 & 0 & 1
\end{array}\right),
\end{equation}
where $s_{ij} = \sin \theta_{ij}$ and $c_{ij} = \cos \theta_{ij}$ are defined by the mixing angles $\theta_{12}$, $\theta_{13}$ and $\theta_{23}$, and $\delta_{CP}$ is the Dirac {\em CP} phase. In addition to the PMNS matrix, the oscillation probabilities depend on the three neutrino mass states $m_1$, $m_2$ and $m_3$, which can be arranged in either normal ordering, $m_3 > m_2 > m_1$, or inverted ordering, $m_2 > m_1 > m_3$. The oscillation frequencies are defined by the mass-square differences {\color{black}$\Delta m_{21}^2 \equiv m_2^2 - m_1^2$} and $\Delta m_{31}^2 \equiv m_3^2 - m_1^2$. Together these parameters are known as the standard oscillation parameters.

Neutrino oscillations become subject to matter effects when neutrinos traverse in a medium.  This phenomenon is known as the Mikheyev-Smirnov-Wolfenstein (MSW) effect~\cite{Wolfenstein:1977ue,Mikheev:1986gs}. The effective Hamiltonian responsible for the neutrino propagation in matter can be written as:
\begin{equation}
\label{eq:HSI}
H = \frac{1}{2E_{\nu}}\left[U
\left(
\begin{array}{ccc}
0 \,\,\, & 0 & 0 \\
0 \,\,\, & \Delta m_{21}^2 & 0\\
0 \,\,\, & 0 & \Delta m_{31}^2 
\end{array}
\right) U^{\dagger}
+ A_{CC}
\left(
\begin{array}{ccc}
1 & 0 & 0 \\
0 & 0 & 0 \\
0 & 0 & 0
\end{array}
\right)
\right],
\end{equation}
where $E_\nu$ is the neutrino energy, $U$ is the PMNS matrix and {\color{black}$A_{CC}$ is the charged current matter potential arising from exchanges of $W$ bosons with the medium}. For neutrinos, the matter potential is defined as $A_{CC} = \sqrt{2} G_F N_e$, with $G_F$ denoting the Fermi constant and $N_e$ the electron number density in the medium. For antineutrinos, the matter potential is obtained from the transformation $A_{CC} \rightarrow -A_{CC}$.

{\color{black}
The oscillation probabilities can be calculated from the effective Hamiltonian\,(\ref{eq:HSI}) as $P_{\nu_\ell \rightarrow \nu_{\ell'}} = \langle \nu_{\ell'}| e^{-iHL}|\nu_\ell\rangle$. In the leading order, when $\sin \theta_{13}$ and $\alpha = \Delta m_{21}^2 / \Delta m_{31}^2$ are taken to be small, neutrino oscillations corresponding to $\nu_e \rightarrow \nu_\mu$ and $\nu_e \rightarrow \nu_\tau$ channels can be approximated with~\cite{Kopp:2008ds}:
\begin{equation}
    \label{eq:nue_numu_si}
    \begin{split}
    P_{\nu_e \rightarrow \nu_\mu}^{\rm SO} &= \sin^2 2\theta_{13} \sin^2 \theta_{23} \frac{\sin^2((1-\hat{A})\Delta)}{(1-\hat{A})^2} \\
    &+ \alpha \sin 2\theta_{13} \sin 2\theta_{12} \sin 2\theta_{23} \cos(\delta_{CP} - \Delta) \frac{\sin(\hat{A}\Delta)}{\hat{A}} \frac{\sin(1-\hat{A}\Delta)}{1-\hat{A}},
    \end{split}
\end{equation}

\begin{equation}
    \label{eq:nue_nutau_si}
    \begin{split}
    P_{\nu_e \rightarrow \nu_\tau}^{\rm SO} &= \sin^2 2\theta_{13} \cos^2 \theta_{23} \frac{\sin^2((1-\hat{A})\Delta)}{(1-\hat{A})^2} \\
    &- \alpha \sin 2\theta_{13} \sin 2\theta_{12} \sin 2\theta_{23} \cos(\delta_{CP} - \Delta) \frac{\sin(\hat{A}\Delta)}{\hat{A}} \frac{\sin(1-\hat{A}\Delta)}{1-\hat{A}},
    \end{split}
\end{equation}
where the superscript SO stands for standard oscillations, $\hat{A} = A_{CC}/\Delta m_{31}^2$, $\Delta = L \Delta m_{31}^2 / 4 E_\nu$, and $L$ denotes the distance neutrinos have traversed. As one can see from the analytic expressions presented in equations\,(\ref{eq:nue_numu_si}) and (\ref{eq:nue_nutau_si}), the appearance channels corresponding to $\nu_e \rightarrow \nu_\mu$ and $\nu_e \rightarrow \nu_\tau$  oscillations are both sensitive to the standard oscillation parameters $\theta_{23}$, $\delta_{CP}$ as well as $\Delta m_{31}^2$.
}

\subsection{Oscillations with a non-unitary mixing matrix}
\label{sec:prob2}

As the existence of the neutrino oscillations itself is a direct evidence of physics beyond the Standard Model, it is worthwhile to consider whether there could be non-standard physics present in the neutrino mixing itself. One popular form of new physics manifests itself as non-unitarity in the neutrino mixing matrix, as is the case with theoretical models featuring Type-I and Type-III Seesaw mechanism~\cite{Schechter:1980gr}.

When one studies a model where the neutrino mixing matrix might not be unitary, the mixing between the active and sterile states can be presented in the block form~\cite{Hettmansperger:2011bt}:
\begin{equation}
\label{eq:nu}
    {\cal U} = \left(
\begin{array}{cc}
N & V \\
S & T
\end{array}
\right),
\end{equation}
where ${\cal U}$ is the unitary 3+{\em n}$\times$3+{\em n} matrix describing the mixing between the three active neutrinos and {\em n} sterile neutrinos. The mixing between the active states is described by $N$, whereas active-sterile, sterile-active and sterile-sterile mixing are represented by $V$, $S$ and $T$, respectively. In this particular scenario the matrix $N$ may not necessarily be unitary, which has several implications on the fit values of the standard oscillation parameters in the global neutrino oscillation data. In absence of $\nu_\mu \rightarrow \nu_\tau$ oscillations in the conventional superbeam experiments, for example, the matrix element $|U_{\tau 3}|^2$ is only present as a sub-leading term in $\nu_\mu \rightarrow \nu_\mu$ oscillations, whereas the information on $|U_{\tau 1}|^2$ and $|U_{\tau 2}|^2$ is lost in such scenario. One therefore needs either excellent statistics from the $\nu_\mu \rightarrow \nu_\mu$ channel or access to $\nu_\mu \rightarrow \nu_\tau$ channel to recover the sensitivities to $|U_{\tau 1}|$, $|U_{\tau 2}|$ and $|U_{\tau 3}|$ in presence of non-unitary mixing.

{\color{black}
Non-unitary mixing can also affect neutrino propagation in matter. The non-unitarity of $N$ alters the effective Hamiltonian describing three-neutrino oscillations, where the neutral currents can no longer be ignored:
\begin{equation}
\label{eq:HNU}
H = \frac{1}{2E_{\nu}}\left(
\begin{array}{ccc}
0 & 0 & 0 \\
0 & \Delta m_{21}^2 & 0\\
0 & 0 & \Delta m_{31}^2 
\end{array}
\right) + N^\dagger \left(
\begin{array}{ccc}
A_{CC} - A_{NC} & 0 & 0 \\
0 & - A_{NC} & 0 \\
0 & 0 & - A_{NC}
\end{array}
\right) N.
\end{equation}
Here $A_{NC}$ corresponds to the matter potential arising from neutral current interactions. The neutrino oscillation probabilities are hence computed from $P_{\nu_{\ell} \rightarrow \nu_{\ell'}} = |\langle \nu_{\ell'}|e^{-i HL}|\nu_\ell\rangle|^2$ where $H$ respects equation\,(\ref{eq:HNU}).
}

There are multiple ways to parameterize the non-unitarity of the mixing between the three active neutrinos. The corresponding non-unitary matrix $N$ can be decomposed into unitary and non-unitary parts~\cite{Escrihuela:2015wra,Escrihuela:2016ube}:
\begin{equation}
\label{eq:nu1}
    N = \left(
\begin{array}{ccc}
\alpha_{11} & 0 & 0 \\
\alpha_{21} & \alpha_{22} & 0 \\
\alpha_{31} & \alpha_{32} & \alpha_{33}
\end{array}
\right) U_{\rm PMNS},
\end{equation}
where $\alpha_{11}$, $\alpha_{22}$ and $\alpha_{33}$ are real parameters and retain values close to unity, while the off-diagonal parameters $\alpha_{21}$, $\alpha_{31}$ and $\alpha_{32}$ are small and may attain complex values\footnote{Parameters $\alpha_{i j}$ ($i$, $j =$ 1, 2, 3) can also be expressed in terms of matrix elements: $\alpha_{i i} = 1-\frac{1}{2}\sum_{k}\left|{\cal U}_{k i}\right|^2$ and $\alpha_{i j} = \sum_{k}{\cal U}_{i k} {\cal U}_{j k}^{
*}$. Here $k$ runs over the matrix elements which are not included in $N$.}. Note that $U_{\rm PMNS}$ stands for the PMNS matrix. Present constraints on the non-unitarity parameters can be found in~\cite{Escrihuela:2016ube,Blennow:2016jkn}, for example. 

{\color{black}
Non-unitarity of neutrino mixing can be tested in accelerator neutrino experiments with various oscillation channels. The implications of non-unitarity parameters in neutrino oscillation probabilities have been discussed extensively in the literature, see e.g. Refs.~\cite{Malinsky:2009df,Meloni:2009cg,Rodejohann:2009ve,Miranda:2016wdr,Forero:2021azc,Agarwalla:2021owd}. The probability for the muon neutrino disappearance channel $\nu_\mu \rightarrow \nu_\mu$ is known to have sensitivity to parameters $\alpha_{22}$, $\alpha_{21}$ and $\alpha_{32}$, whereas the probability for $\nu_e \rightarrow \nu_e$ is mainly affected by $\alpha_{11}$. The probabilities associated with the appearance channels $\nu_e \rightarrow \nu_\mu$ and $\nu_e \rightarrow \nu_\tau$ on the other hand can be used to probe the third row parameters $\alpha_{31}$ and $\alpha_{33}$.

The analytic expressions to the oscillation probabilities $P_{\nu_e \rightarrow \nu_\mu}$ and $P_{\nu_e \rightarrow \nu_\tau}$ can be derived in the same manner as what is done in the case of standard oscillations in section\,(\ref{sec:prob1}). Expanding $P_{\nu_e \rightarrow \nu_\mu}$ and $P_{\nu_e \rightarrow \nu_\tau}$ in powers of $\alpha$ and $\sin \theta_{13}$ in presence of non-unitary mixing gives rise to the following expressions:
\begin{equation}
    \label{eq:nue_numu_nu}
    \begin{split}
    P_{\nu_e \rightarrow \nu_\mu}^{\rm NU} &= P_{\nu_e \rightarrow \nu_\mu}^{\rm SO} 
        + |\alpha_{31}|\alpha_{33} \sin 2\theta_{13} \sin 2\theta_{23} \cos(\delta_{CP} - \varphi_{31} - \Delta) \frac{\sin(\hat{A}\Delta)}{\hat{A}} \frac{\sin(1-\hat{A}\Delta)}{1-\hat{A}} \\
        &- 2 |\alpha_{31}|\alpha_{33} \sin 2\theta_{13} \sin^2 \theta_{23} \cos(\delta_{CP} - \varphi_{31}) \frac{\hat{A} \sin^2((1-\hat{A})\Delta)}{(1-\hat{A})^2},
    \end{split}    
\end{equation}

\begin{equation}
    \label{eq:nue_nutau_nu}
    \begin{split}
    P_{\nu_e \rightarrow \nu_\tau}^{\rm NU} &= P_{\nu_e \rightarrow \nu_\tau}^{\rm SO} 
        - |\alpha_{31}|\alpha_{33} \sin 2\theta_{13} \sin 2\theta_{23} \cos(\delta_{CP} - \varphi_{31} - \Delta) \frac{\sin(\hat{A}\Delta)}{\hat{A}} \frac{\sin(1-\hat{A}\Delta)}{1-\hat{A}} \\
        &- 2 |\alpha_{31}|\alpha_{33} \sin 2\theta_{13} \cos^2 \theta_{23} \cos(\delta_{CP} - \varphi_{31}) \frac{\hat{A} \sin^2((1-\hat{A})\Delta)}{(1-\hat{A})^2},
    \end{split}
\end{equation}
where $P_{\nu_e \rightarrow \nu_\mu}^{\rm SO}$ and $P_{\nu_e \rightarrow \nu_\tau}^{\rm SO}$ denote the standard oscillation probabilities in the corresponding channels. The analytic expressions given in equations\,(\ref{eq:nue_numu_nu}) and (\ref{eq:nue_nutau_nu}) present the oscillation probabilities under non-unitary neutrino mixing up to first order of $\alpha$. The expressions show that the probabilities are driven by parameters $\alpha_{31}$ and $\alpha_{33}$. It is noteworthy that the second and third term in $P_{\nu_e \rightarrow \nu_\mu}^{\rm NU}$ have opposite signs, whereas the corresponding terms in $P_{\nu_e \rightarrow \nu_\tau}^{\rm NU}$ have the same sign. Depending on the values of $\delta_{CP}$ and $\varphi_{31}$, this structure could lead to cancellations in one oscillation channel and enhancements in the other. One can therefore expect complementarity from the two appearance channels.
}

\subsection{Non-standard neutrino interactions}
\label{sec:prob3}

In addition to the standard precision tests on three-neutrino mixing, future neutrino experiments are also able to look for new physics in neutrino interactions. In the quantum mechanical description of neutrino oscillations, this form of new physics is typically described with the non-standard interaction (NSI) parameters, which parameterize new interactions in terms of the Fermi coupling constant $G_F$. The incoherent production and detection of neutrinos give rise to NSI in the source and detection, which are expressed by the source and detection NSI parameters $\epsilon^{s}_{\ell \ell'}$ and $\epsilon^{d}_{\ell \ell'}$, where $\ell$, $\ell' = e$, $\mu$ and $\tau$ stand for the neutrino flavour~\cite{Grossman:1995wx,Ohlsson:2008gx,Farzan:2017xzy}:
\begin{eqnarray}\label{eq:cc-nsi}
|\nu_\ell^s\rangle = \frac{(1+\epsilon^s)_{\ell \alpha}}{N^s_\ell}|\nu_\alpha\rangle,\ \langle\nu_{\ell'}^d| = \langle\nu_\alpha| \frac{(1+\epsilon^d)_{\alpha \ell'}}{N^d_{\ell'}}.
\end{eqnarray}

In the expressions shown in equation\,(\ref{eq:cc-nsi}), the flavour states representing the neutrino in the production and detection, $|\nu_\ell^s\rangle$ and $\langle\nu_{\ell'}^d|$ respectively, are related to the eigenstates via the source and detection NSI parameters with the normalization factors $N_{\ell}^s=\sqrt{[(1+\epsilon^s)(1+\epsilon^{s\dagger})]_{\ell \ell}}$ and $N_{\ell'}^d=\sqrt{[(1+\epsilon^{d{\dagger}})(1+\epsilon^d)]_{\ell' \ell'}}$. Neutrino oscillations are furthermore affected by NSI effects in the propagation. In the effective Hamiltonian, the standard matter effects are complemented with matter NSI parameters $\epsilon^m_{\ell \ell'}$ as follows:
\begin{equation}
\label{eq:HNSI}
H = \frac{1}{2E_{\nu}}\left[U
\left(
\begin{array}{ccc}
0 \,\,\, & 0 & 0 \\
0 \,\,\, & \Delta m_{21}^2 & 0\\
0 \,\,\, & 0 & \Delta m_{31}^2 
\end{array}
\right) U^{\dagger}
+ A
\left(
\begin{array}{ccc}
1+\varepsilon_{ee}^m & \epsilon_{e\mu}^m & \,\,\, \epsilon_{e\tau}^m \\
\epsilon_{e\mu}^{m*} & \epsilon_{\mu\mu}^m & \,\,\, \epsilon_{\mu\tau}^m \\
\epsilon_{e\tau}^{m*} & \epsilon_{\mu\tau}^{m*} & \,\,\, \epsilon_{\tau\tau}^m
\end{array}
\right)
\right],
\end{equation}
where the diagonal elements of the $\epsilon^m$ matrix are real, whereas the off-diagonal parameters can be complex numbers. The probability for a neutrino of flavour $\ell$ to oscillate into a neutrino of flavour $\ell'$ is then given by $P_{\ell \ell'}=|\langle \nu_{\ell'}^d|e^{-i HL}|\nu_\ell^s\rangle|^2$. 

Source and detection NSI parameters are strictly constrained by experimental data from short-baseline experiments~\cite{Biggio:2009nt}, whereas bounds on matter NSI parameters are relatively large~\cite{Blennow:2016etl}. Future experiments with long baseline lengths are most suitable to study NSI effects in neutrino propagation. Especially tau neutrino physics is known to be beneficial in the measurement of $\epsilon^{m}_{\ell \ell'}$ parameters in the third row of equation\,(\ref{eq:HNSI})~\cite{Autiero:2003fu,Kopp:2008ds}.

{\color{black}
The phenomenological consequences of the NSI parameters have been discussed in the literature, see e.g. Refs.~\cite{Ribeiro:2007ud,Kopp:2008ds,Meloni:2009cg,Blennow:2016jkn}. The oscillation probabilities in $\nu_e \rightarrow \nu_\mu$ and $\nu_e \rightarrow \nu_\tau$ channels can be expressed with the following analytical expressions~\cite{Kopp:2008ds}:
\begin{equation}
    \label{eq:nue_numu_nsi}
    \begin{split}
    P_{\nu_e \rightarrow \nu_\mu}^{\rm NSI} &= P_{\nu_e \rightarrow \nu_\mu}^{\rm SO} 
        - 2 |\epsilon_{e \tau}^m| \sin 2\theta_{13} \sin 2\theta_{23} \cos(\delta_{CP} + \phi_{e \tau}^m - \Delta) \frac{\sin(\hat{A}\Delta)}{\hat{A}} \frac{\sin(1-\hat{A}\Delta)}{1-\hat{A}} \\
        &+ 4 |\epsilon_{e \tau}^m| \sin 2\theta_{13} \sin^2 \theta_{23} \cos(\delta_{CP} + \phi_{e \tau}^m) \frac{\hat{A} \sin^2((1-\hat{A})\Delta)}{(1-\hat{A})^2},
    \end{split}    
\end{equation}

\begin{equation}
    \label{eq:nue_nutau_nsi}
    \begin{split}
    P_{\nu_e \rightarrow \nu_\tau}^{\rm NSI} &= P_{\nu_e \rightarrow \nu_\tau}^{\rm SO} 
        + 2 |\epsilon_{e \tau}^m| \sin 2\theta_{13} \sin 2\theta_{23} \cos(\delta_{CP} + \phi_{e \tau}^m - \Delta) \frac{\sin(\hat{A}\Delta)}{\hat{A}} \frac{\sin(1-\hat{A}\Delta)}{1-\hat{A}} \\
        &+ 4 |\epsilon_{e \tau}^m| \sin 2\theta_{13} \cos^2 \theta_{23} \cos(\delta_{CP} + \phi_{e \tau}^m) \frac{\hat{A} \sin^2((1-\hat{A})\Delta)}{(1-\hat{A})^2},
    \end{split}
\end{equation}
where the probabilities have been expressed up to the first order of $\alpha$. We have neglected the NSI parameters corresponding to the source and detection processes, as the present constraints to both $\epsilon^s$ and $\epsilon^d$ are more than an order of magnitude lower than the ones associated with matter NSI effects~\cite{Blennow:2016etl}. As one can see, the appearance probabilities $P_{\nu_e \rightarrow \nu_\mu}^{\rm NSI}$ and $P_{\nu_e \rightarrow \nu_\tau}^{\rm NSI}$ are influenced by the matter NSI parameter $\epsilon_{e \tau}^m$.

It is interesting to note that the oscillation probabilities derived for the matter NSI case are similar to the ones that were obtained for non-unitary neutrino mixing. A comparison between equations\,(\ref{eq:nue_numu_nsi}) and (\ref{eq:nue_nutau_nsi}) and equations\,(\ref{eq:nue_numu_nu}) and (\ref{eq:nue_nutau_nu}) shows that $|\epsilon_{e \tau}^m|$ makes a stronger contribution to the appearance probabilities in $\nu_e \rightarrow \nu_\mu$ and $\nu_e \rightarrow \nu_\tau$ channels than $|\alpha_{31}|$, as the leading terms in the corresponding probabilities are different by a factor of 2. It can therefore be expected that experiments with sensitivities to channels $\nu_e \rightarrow \nu_\mu$ and $\nu_e \rightarrow \nu_\tau$ can lead to more stringent bounds on the NSI parameter $\epsilon_{e \tau}^m$ than on the non-unitarity parameter $\alpha_{31}$.
}

{\color{black}We finally note that long-baseline neutrino experiments can display significant sensitivities to new physics such as non-unitary mixing and non-standard interactions with matter. As we have shown in this section, the oscillation probabilities that are relevant for long baseline lengths can be used to probe non-unitarity and NSI parameters. It has furthermore been shown in the literacy that near detectors can also be very sensitive to such effects via the zero-distance effect~\cite{Ohlsson:2012kf,Miranda:2016wdr}. Future studies are needed to understand the full interplay of the near and far detectors.}

\section{Overview of the accelerator and underground laboratories in China}
\label{sec:labsites}

There are a number of large-scale experimental research laboratories that are currently underway in China. In this section, we present a survey on seven notable laboratories that could be considered as candidates for a neutrino source or detector site.

\subsection{Accelerator laboratories}

At the moment, there are five different laboratories and institutes with the capability to host a proton accelerator center. Three of them, CSNS, CiADS and CAS-IMP, are already in operation and underway to reach their full potential. Two other laboratories, Nanjing University and SPPC, are going through the planning phase.

\subsubsection*{CSNS}
\noindent China Spallation Neutrino Source (CSNS) is an accelerator-based neutron source commissioned to conduct experiments with neutrons~\cite{CSNS2006}. It is one of the largest on-ground research facilities in China and its primary purpose is to develop novel methods for material characterization using neutron scattering techniques. {\color{black}CSNS} is stationed in Dongguan, Guangdong, and includes linear proton accelerator, rapid circling synchrotron and a target station with three neutron instruments. The accelerator facility in CSNS is home to a 1.6~GeV proton driver with 100~kW beam power in its first stage (CSNS~I), which will be upgraded to 500~kW for its second stage (CSNS~II). The CSNS accelerator facility has ultimately the prospects to reach 4~MW beam power and 128~GeV proton energy, when a post-acceleration system is added (CSNS+). In addition to its primary goal of serving neutron science, the CSNS accelerator laboratory also provides opportunities to conduct fundamental research. Applications to use the post-accelerated protons of the CSNS+ phase have been studied in a neutrino superbeam experiment~\cite{Yang:2013afa}. In such case, extracting only roughly 10\% of the CSNS protons from the neutron source would allow to generate 4~MW proton beam of 128~GeV energy, which could be used to produce neutrinos via pion decay. It is also possible to consider further prospects in producing neutrinos via muon beams, with the Experimental Muon Source (EMuS) currently in place in CSNS. 

CSNS could make an attractive location for a future accelerator neutrino facility, with its location of 162~km from JUNO and 1329~km from CJPL. It could therefore be used in medium-baseline oscillation experiments as well as in long-baseline neutrino experiments.

\subsubsection*{CiADS}
\noindent China initiative for Accelerator Driven System (CiADS) is a strategic plan to solve the nuclear waste and resource problems concerning the future nuclear power industry in China~\cite{CiADS2012}. The initiative entails a long-term experimental program where an accelerator-driven sub-critical demonstration facility is built and operated over a staged 20-year plan. The first stage of the program involving a proton beam of 100~kW power and 1.5~GeV energy is presently operational. The facility delivers a 10~mA beam and runs entirely continuous-wave mode. The second stage, where the system is to be upgraded to 500~kW beam power, is already funded. At the end {\color{black}of} its program, the facility is expected to reach proton beam energies as high as 15~MW in continuous-wave mode. The technology developed in the CiADS program and even the facility itself can also be used in a neutrino physics program. The idea of building a neutrino experiment has previously been {\color{black}reported} in the MOMENT proposal~\cite{Cao:2014bea}.

CiADS is located in the city of Huizhou in Guangdong, approximately 221~km from the JUNO detector site and 1389~km from CJPL. Whereas the distance to JUNO is suitable for a low-energy neutrino beam facility like MOMENT, the longer baseline involved in the CJPL site could be used to study higher neutrino energies, including the beam energies that are currently being planned for DUNE.

\subsubsection*{CAS-IMP}
\noindent Institute of Modern Physics of the Chinese Academy of Sciences (CAS-IMP) is a major research institute located in the city of Lanzhou in central China. The institute was founded in 1957 and it has developed into the most important Chinese research center focusing on heavy ion physics. The institute has a long-running tradition in accelerator physics, employing many experts in heavy ion physics while operating the Heavy Ion Research Facility in Lanzhou~\cite{Mao:2020rlb}. CAS-IMP is also in charge of establishing the next major Chinese heavy ion physics laboratory High Intensity Heavy-ion Accelerator~\cite{Yang:2013yeb} in the southern province of Guangdong, not far from the CiADS laboratory. Whereas the accelerator facility in Lanzhou is able to propel ions up to 1~GeV, the accelerator in Huizhou operates with proton energies up to 800~MeV.

The CAS-IMP site in Lanzhou is located approximately 894~km from CJPL and 1759~km from JUNO, making it a viable candidate for long-baseline neutrino oscillation physics. In this work, we consider CAS-IMP in Lanzhou as one of the potential sites for an accelerator neutrino facility in a future experiment.

\subsubsection*{Nanjing}
\noindent Nanjing University is one of the major public universities in China. Several years ago, the university initiated a new technology development program in the field of high-energy charged particle beam applications and fundamental sciences~\cite{Sun:2017}. A key part of the program is the Proton Linear Accelerator Institute where plans to build a high-current proton linear accelerator were recently unveiled. The proton accelerator will serve for a variety of purposes, ranging from radio-isotopes and medical applications to nuclear physics and material sciences. The original plan for the accelerator aims at energy range of [10, 1000]~MeV with a 26~mA current. The operation module may be chosen from either the continuous-wave or pulsed beam technique, with operation frequencies of 403~MHz and 806~MHz. In future upgrades the accelerator laboratory could potentially host additional programs dedicated to fundamental sciences. 

The Proton Linear Accelerator Institute of Nanjing University is suitable for an accelerator neutrino facility in a long-baseline oscillation experiment, being located roughly 1261~km from JUNO and 1693~km from CJPL.

\subsubsection*{SPPC}
\noindent Super Proton-Proton Collider (SPPC) is a future high-energy collider and the second stage of Circular Electron-Positron Collider (CEPC) currently in planning in China~\cite{CEPC-SPPCStudyGroup:2015csa,CEPC-SPPCStudyGroup:2015esa}. The experimental program of SPPC includes the measurements on the Higgs couplings, with an aim to measure several rare decay processes and probing the Higgs self-coupling and $Htt$ coupling. Knowledge on these coupling strengths is considered to be crucial in understanding the form of the Higgs potential, and a direct measurement on the $HHH$ coupling could help to understand whether the electroweak phase transition is of the first order or second order. 

The SPPC complex consists of a large collider ring of 100~km circumference, with two interaction points and a nominal luminosity of 1.0$\times$10$^{35}$~cm$^{-2}$s$^{-1}$ in each. The collider is expected to reach 75~TeV center of mass energy in its initial stage, with a prospect to ultimately reach 125--150~TeV~\footnote{High-luminosity upgrades and non-collider physics can also be considered for SPPC. In a recent study~\cite{Canbay:2017rbg}, for example, the SPPC accelerator was studied as a host for a lepton-proton collider.}. SPPC also includes a four-stage injector chain, which comprises a proton linac and three synchrotron rings. The injector chain delivers 2.1~TeV proton beam to the SPPC ring. The duty cycle of the injector chain also allows to consider non-collider physics programs, with the potential to divert high-energy protons at 3.2~MW average beam power~\cite{CEPC-SPPCStudyGroup:2015csa}.

In the present work, we investigate the feasibility of realizing a future neutrino source in the SPPC accelerator facility near Beijing. The neutrino source would be located about 1736~km from CJPL and 1871~km from JUNO, respectively.

\subsection{Underground laboratories}

There are presently two major underground laboratories in China. While CJPL is already in place and ready for extensions, the civil construction in JUNO is underway.

\subsubsection*{CJPL}
\noindent China JinPing Laboratory (CJPL) is one of the two underground laboratories currently in operation in China, built under the Jinping mountain in Sichuan. Located more than 2400 meters underground, CJPL is currently the deepest underground laboratory in the world. It has relatively low cosmic muon flux and muon-induced background. The laboratory is also far from the major nuclear power plants, giving it low reactor neutrino background. CJPL is therefore {\color{black}an} optimal location to study the physics of solar neutrinos, dark matter and supernova neutrinos. The laboratory currently hosts two dark matter experiments, PandaX and CDEX, and there are aspirations to build a neutrino detector {\color{black}to study low-energy neutrinos of solar and astrophysical origin~\cite{Jinping:2016iiq}}.

\subsubsection*{JUNO}
\noindent Jiangmen Underground Neutrino Observatory (JUNO) is a reactor neutrino experiment designed to study reactor neutrinos from the Taishan and Yangjiang nuclear power plants in the province of Guangdong in China. The main goal of JUNO is to determine the neutrino mass ordering by at least 3\,$\sigma$ CL. The experiment hosts an underground laboratory located 53~km from both Taishan and Yangjiang. The civil construction began in 2015 and the experiment is expected to be operational in 2023\footnote{Owing to several setbacks, it is not yet known when JUNO will begin to take data.}. The laboratory provides the facilities for a {\color{black}highly} transparent liquid scintillator detector of 20~kton fiducial mass and it is planned to be operational for {\color{black}20} years. {\color{black}The central detector filled with liquid scintillator in JUNO is located underground, surrounded by a water pool as a Cherenkov detector and covered by the plastic scintillator as a top tracker. This leads to the relatively low backgrounds from cosmic ray muons and makes JUNO suitable for low-energy neutrino physics.
}

\subsection*{Choosing location for accelerator and detector facilities}

The compatibility of the source and detector locations can be evaluated with the position on the oscillation maximum. Oscillation maximum defines the neutrino energy for which the conversion $\nu_{\ell} \rightarrow \nu_{\ell'}$, $\ell \neq \ell'$ occurs at its local maximum. Focusing the neutrino beam on one or more oscillation maxima is therefore preferred in an oscillation experiment. In accelerator neutrino experiments with medium and long baseline lengths, the oscillations in the channels $\nu_\mu \rightarrow \nu_e$, $\nu_e \rightarrow \nu_\mu$ and $\nu_\mu \rightarrow \nu_\tau$ reach their maximum when the following condition is satisfied:
\begin{equation}
    \label{eq:oscmax}
    \frac{L \Delta m_{31}^2}{4 E_\nu} = \frac{(2n + 1) \cdot \pi}{2},
\end{equation}
where $n =$ 0 for the first maximum and $n =$ 1 for the second maximum\footnote{It should be noted that equation\,(\ref{eq:oscmax}) defines oscillation maximum for $\Delta m_{31}^2$--driven neutrino oscillations. In a similar manner, one may derive the relation for $\Delta m_{21}^2$--driven oscillation mode. An important yet largely unexplored topic is the interference of $\Delta m_{21}^2$ and $\Delta m_{31}^2$--driven frequencies in neutrino oscillation experiments and their effect on the oscillation maximum.}. Using the geographical locations of the laboratories discussed in this section, we present the baseline lengths as well as the approximate neutrino energies for the first and second oscillation maxima in table\,\ref{tab:labsites2}. {\color{black}The available baseline lengths are calculated from the geographical coordinates and range from 162~km to 1871~km.} In the following sections, we investigate physics prospects of the laboratory combinations mentioned in this section\footnote{Two proposals for accelerator facilities have already been produced. Post-accelerated protons from CSNS~\cite{Yang:2013afa} and CiADS-type accelerator~\cite{Cao:2014bea} have been considered as potential neutrino sources utilizing pion and muon decays, respectively.}.

{\color{black}It is now appropriate to comment on the timelines that have been expected for the various accelerator and detector programs discussed in this section. CSNS concluded its Phase-I installation in 2018 and the very early user instruments were finished in early 2021. CiADS began its first two stages in 2011 and 2016 respectively, whereas the third stage is expected to begin operations in 2030. SPPC will follow after CEPC, which in turn is projected to start taking data in 2034--2036. At the same time, JUNO and CJPL can be expected to be available for upgrades after about 2030s, when the present experimental programs have been completed. The schedules of the potential CAS-IMP and Nanjing University are still unclear. It could be, therefore, predicted that any future accelerator neutrino program could take place after the present accelerator programs are complete in 2040--2060.}
\begin{table}[!t]
\caption{\label{tab:labsites2}The distances between the planned and existing accelerator and underground laboratories in China. {\color{black}Geographical coordinates with latitude and longitude in degrees are provided in parentheses.} Also provided are the approximate energies for the first and second oscillation maxima, assuming $\Delta m_{31}^2 \simeq$ 2.517$\times$10$^{-3}$~eV$^2$.}
\begin{center}
\resizebox{\linewidth}{!}{
\begin{tabular}{ |c|c|c|c|c|c|c| } 
 \hline
 \multirow{2}{*}{Accelerator facility} & \multicolumn{3}{ c| }{JUNO~(22.12$^\circ$,~112.51$^\circ$)} & \multicolumn{3}{ c| }{CJPL~(28.15$^\circ$,~101.71$^\circ$)} \\
 \cline{2-7}
 & Baseline & 1$^{\rm st}$ maximum & 2$^{\rm nd}$ maximum & Baseline & 1$^{\rm st}$ maximum & 2$^{\rm nd}$ maximum \\
 \hline
 CAS-IMP~(36.05$^\circ$,~103.68$^\circ$) & {\color{black}1759~km} & {\color{black}3.6~GeV} & {\color{black}1.2~GeV} & {\color{black}894~km} & {\color{black}1.8~GeV} & {\color{black}600~MeV} \\ \hline
 CiADS~(23.08$^\circ$,~114.40$^\circ$) & {\color{black}221~km} & {\color{black}450~MeV} & {\color{black}150~MeV} & {\color{black}1389~km} & {\color{black}2.8~GeV} & {\color{black}940~MeV} \\ \hline
 CSNS~(23.05$^\circ$,~113.73$^\circ$) & {\color{black}162~km} & {\color{black}330~MeV} & {\color{black}110~MeV} & {\color{black}1329~km} & {\color{black}2.7~GeV} & {\color{black}900~MeV} \\ \hline
 Nanjing~(32.05$^\circ$,~118.78$^\circ$) & {\color{black}1261~km} & {\color{black}2.6~GeV} & {\color{black}850~MeV} & {\color{black}1693~km} & {\color{black}3.4~GeV} & {\color{black}1.1~GeV} \\ \hline
 SPPC~(39.93$^\circ$,~116.40$^\circ$) & {\color{black}1871~km} & {\color{black}3.8~GeV} & {\color{black}1.3~GeV} & {\color{black}1736~km} & {\color{black}3.5~GeV} & {\color{black}1.2~GeV} \\
 \hline
\end{tabular}}
\end{center}
\end{table}

\section{Description of the experimental configuration}
\label{sec:conf}

In order to understand which of the available experimental sites provide favourable conditions for a future accelerator-based neutrino oscillation experiment, we perform a simulation study of hypothetical neutrino oscillation experiment utilizing muon-decay and beta beam technologies. In this section, we define the key parameters of the experimental configurations that are considered in our study. The general setup for the simulated neutrino experiment is discussed in section~\ref{sec:conf1}. Methods for the statistical analysis are summarized in section~\ref{sec:conf2}.

\subsection{General configuration for an accelerator neutrino experiment in China}
\label{sec:conf1}

As the numerical part of this work, we investigate the prospects of the neutrino experiment configurations that could be established using accelerator and underground laboratories in China. In this subsection, we briefly review the available neutrino beam and detector options which could be considered for such a neutrino oscillation experiment. We furthermore discuss what could be a concrete proposal for a future accelerator-based neutrino oscillation experiment based in China.

Accelerator neutrino experiments operating over long baseline lengths can be divided into three stages: neutrino production, propagation and detection. Each of these stages play an important role in determining the success of any future experiment. The neutrino production method could be based on the pion decay-in-flight, muon decay-in-flight or ion decay-in-flight neutrino beam technologies~\cite{Charitonidis:2021qfm}. Neutrino detectors on the other hand could be constructed using any of the presently established technologies, which include Water Cherenkov (W.C.), Magnetized Iron (M.I.) and Liquid Scintillator (L.Sc.) neutrino detector concepts. Hybrid detectors combining one or more detection methods could also be considered. {Successful examples of neutrino detectors utilizing the aforementioned techniques are found in NO$\nu$A~\cite{NOvA:2019cyt}, T2K~\cite{T2K:2019bcf} and MINOS~\cite{MINOS:2006foh}.} There are also new types of detector technologies currently under investigation, the most notably example being the Liquid Argon Time Projection Chamber (LArTPC)~\cite{Majumdar:2021llu} currently being developed for DUNE~\cite{Abi:2020loh}. Emerging neutrino detector technologies are also under active R\&D in various collaborations, including projects such as the W.C. and L.Sc. hybrid THEIA~\cite{Gann:2015fba} and opaque detector concept LiquidO~\cite{Cabrera:2019kxi}. As the potential accelerator neutrino oscillation experiment in China would presumably take place in the not-so-near future, any of the beam and detector technologies mentioned above could be considered for its experimental setup.

One of the key questions in the development of future neutrino detectors is their capability to detect tau neutrinos. Tau neutrino appearance channels such as $\nu_\mu \rightarrow \nu_\tau$ and $\nu_e \rightarrow \nu_\tau$ are reported to have a notable potential to increase the sensitivity to many physics-driven goals, such as the precision measurements on standard oscillation parameters, testing unitarity of PMNS matrix as well as searching non-standard effects in neutrino interactions~\cite{DeGouvea:2019kea}. Prospects of using $\nu_\tau$ events to improve prospects to new physics have previously been studied in experiments like OPERA~\cite{Meloni:2019pse} and DUNE~\cite{DeGouvea:2019kea,Abi:2020kei}. The main challenges related to $\nu_\tau$ detection are the relatively high threshold, around 3.5\,GeV, as well as short $\tau$ lifetime. {\color{black}Successful tau neutrino detection furthermore requires advances in detector technology, as hadronic decays of $\tau$ can easily dilute the sensitivity to tau neutrinos.} If these challenges can be addressed, the inclusion of the silver channel $\nu_e \rightarrow \nu_\tau$ could lead to significant improvements in the sensitivity to new physics in neutrino experiments driven by muon decay~\cite{Kopp:2008ds}. The prospects of studying silver channel in beta beam experiments on the other hand have not yet been studied in detail.

  \begin{figure}[!t]
        \center{\includegraphics[width=\textwidth]
        {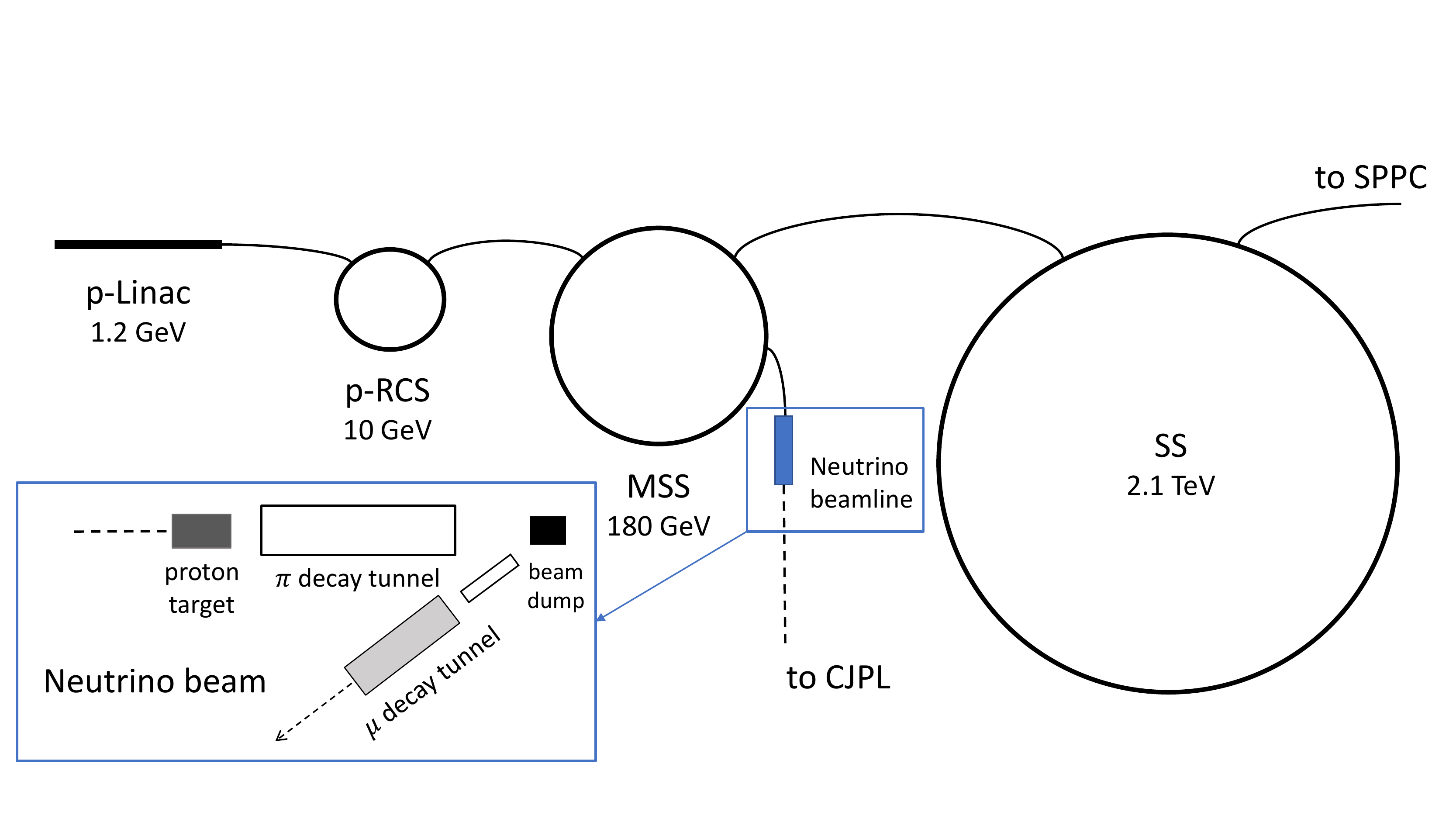}}
        \caption{\label{fig:sppc_inj} A schematic illustration of the SPPC injector chain complemented with a neutrino production beamline. The injector chain consists of four stages, three of which could be used for proton extraction for a neutrino beam. The accelerated proton energies at the end of each stage is 1.2~GeV, 10~GeV, 180~GeV and 2.1~TeV, respectively. The MSS ring appears the most suitable for a neutrino program with its 180~GeV proton beam energy and opportunity to divert 3.2~MW to non-collider programs.}
      \end{figure}

In the present work, we consider the accelerator neutrino beams based on the muon decay-in-flight and beta beam techniques for the future accelerator neutrino experiment. Neutrino beams generated by muon decay have been investigated in great detail in previous studies. The most general study on the prospects and limits of beam technologies based on high-energy muon decay are presented in the International Scope Study for the Neutrino Factory (ISS-NF)~\cite{Choubey:2010zz,Choubey:2011zzq}. Similar setups with lower muon beam energies have previously been considered for shorter baselines in $\nu$STORM and MOMENT proposals~\cite{nuSTORM:2013cqr,Cao:2014bea}. Accelerator neutrino experiments using beta beam technique is a less investigated option due to their relatively high cost. In beta beams, neutrino beams of electron flavour are produced via beta decays of charged ions. This production mechanism is well-known and it is able to produce a very clean neutrino beam with low beam-related backgrounds. Using beta beams in neutrino oscillation experiments have previously been discussed in Refs.~\cite{Huber:2005jk,Coloma:2012wq,Bakhti:2013ama}. In this work, we study the physics potential of a neutrino beam driven by beta decays of accelerated $^{6}$He and $^{18}$Ne isotopes. In contrast to previous studies on beta beam, we also include tau neutrino appearance in the beta beam option.

The most favourable neutrino detector technology for a neutrino beam based on decay-in-flight muons is deemed to be M.I. detector, which provides excellent prospects to study $\nu_e \rightarrow \nu_\mu$ and $\bar{\nu}_\mu \rightarrow \bar{\nu}_e$ oscillations~\cite{Huber:2002mx,Autiero:2003fu,Huber:2006wb}. Such detector setup is often studied together with an emulsion {\color{black}cloud} chamber (E.C.C.) detector, which provides the ability to reconstruct tau neutrinos from $\nu_e \rightarrow \nu_\tau$ oscillations~\cite{Kopp:2008ds}. Beta beams
on the other hand have mainly been studied in large W.C. and L.Sc. detectors, though other technologies can also be considered as well. In the present work, we assume the M.I. and E.C.C. hybrid detector in the muon beam option and L.Sc. detector in the beta beam option, respectively.

In China, a promising location for a future neutrino beam facility is the injector chain of the SPPC accelerator complex~\cite{CEPC-SPPCStudyGroup:2015csa}, which is illustrated in figure\,\ref{fig:sppc_inj}. In the SPPC injector, protons are at first accelerated to 1.2~GeV energy in p-Linac and continue their way to p-RCS and MSS beamlines\footnote{The individual parts of SPPC injector are named as follows: proton linac (p-Linac), rapid  cycling  synchrotron (p-RCS), medium-stage  synchrotron (MSS) and final  stage  synchrotron (SS)~\cite{CEPC-SPPCStudyGroup:2015esa}. The option to use p-RCS and MSS in neutrino beams is first mentioned in Ref.~\cite{Tang:2015qga}.} at 10~GeV and 180~GeV proton energies, respectively. The final stage before injecting to SPPC is the SS accelerator ring, where the protons reach 2.1~TeV energy. A neutrino beam facility could be built either adjacent to p-RCS or MSS rings, which both could divert a 3.2~MW proton beam for a non-collider physics program such as neutrino beams~\cite{CEPC-SPPCStudyGroup:2015csa}. Once the collider begins taking data, it will remain operational for at least 10 years. The p-RCS and MSS rings could also be used for non-collider physics several years prior to the launch of the collider program~\cite{CEPC-SPPCStudyGroup:2015csa}. This gives an opportunity to consider various configurations for a potential neutrino beam.

\begin{table}[!t]
\caption{\label{tab:configuration} Benchmark details of the simulated neutrino oscillation experiments. The studied configuration consists of the muon and beta beam options, where useful parent decays are reported for muons and $^{6}$He ($^{18}$Ne) ions and energy resolution for $\nu_e$/$\bar{\nu}_e$ ($\nu_\mu$/$\bar{\nu}_\mu$) candidates.}
\begin{center}
\begin{tabular}{|c|c|c|}
\hline
\rule{0pt}{3ex}{\bf Parameter} & {\bf Muon beam} & {\bf Beta beam} \\ \hline
\rule{0pt}{3ex}Production method & muon decay-in-flight & ion decay-in-flight \\\hline
\rule{0pt}{3ex}Detection method & hybrid detector & liquid scintillator \\\hline
\rule{0pt}{3ex}Useful parent decays & 2.5$\times$10$^{20}$ year$^{-1}$ & 2.2(5.8)$\times$10$^{18}$year$^{-1}$ \\ \hline
\rule{0pt}{3ex}Detector mass & 50~kton & 50~kton \\ \hline
\rule{0pt}{3ex}Detection threshold & 1~GeV & 0.5~GeV\\ \hline
\rule{0pt}{3ex}Energy resolution & 15\%(15\%)/$E_\nu$ & 6\%(5\%)/$\sqrt{E_\nu}$ \\ \hline
\rule{0pt}{3ex}Energy bins & 45 & 20 \\ \hline
\rule{0pt}{3ex}Running time & 5+5~years & 5+5~years \\ \hline 
\end{tabular}
\end{center}
\end{table}

In this work, we divide the physics-driven study of a future accelerator experiment into two parts. In the first part, we discuss the suitability of the experimental sites described in this work in the muon beam and beta beam configurations. In the second part, we demonstrate the physics potential of an SPPC-based neutrino beamline in such experiment. As the scope of this study is focused on the precision measurement of standard oscillation parameters and physics with tau neutrinos, we call this SPPC-based configuration {\bf PROMPT} ({\em {\bf\em PR}ecisi{\bf\em O}n {\bf\em M}easurements and {\bf\em P}hysics with {\bf\em T}au neutrinos}), for which we adopt a neutrino detector based at CJPL. This gives rise to a baseline length of 1736~km.

\begin{table}[!t]
\caption{\label{tab:channels} The full composition of the signal and background channels considered in the two neutrino beam options described in this work. The simulated configurations for the muon and beta beam options are adapted from the experimental setups described in~\cite{Huber:2002mx,Autiero:2003fu,Huber:2006wb} and~\cite{Huber:2005jk,Coloma:2012wq,Bakhti:2013ama}, respectively. {\color{black}Note that no charge identification is assumed in disappearance channels in the muon beam option.}}
\begin{center}
\resizebox{\linewidth}{!}{%
\begin{tabular}{|llcrr|}
\hline
\rule{0pt}{3ex} {\bf Appearance channels} &    &   &   & {\bf Muon beam option} \\ \hline
\rule{0pt}{3ex} Signal: & \multirow{2}{*}{$\nu_e \rightarrow \nu_\mu$} & & Background: & \multirow{2}{*}{$\bar{\nu}_\mu$ NC and $\bar{\nu}_\mu \rightarrow \bar{\nu}_\mu$ mis-id.} \\ 
\rule{0pt}{3ex}         & & & & \\ 
\rule{0pt}{3ex}         & \multirow{2}{*}{$\bar{\nu}_e \rightarrow \bar{\nu}_\mu$} & & & \multirow{2}{*}{$\bar{\nu}_\mu$ NC and $\nu_\mu \rightarrow \nu_\mu$ mis-id.} \\ 
\rule{0pt}{3ex}         & & & & \\ 
\rule{0pt}{3ex}         & \multirow{2}{*}{$\nu_e \rightarrow \nu_\tau$} & & & $\nu_e \rightarrow \nu_e$, $\nu_e \rightarrow \nu_\mu$, $\bar{\nu}_\mu \rightarrow \bar{\nu}_\mu$, \\ 
\rule{0pt}{3ex}         & & & & $\bar{\nu}_\mu \rightarrow \bar{\nu}_\tau$, $\bar{\nu}_\mu$ NC and $\nu_e$ NC \\ \hline
\rule{0pt}{3ex} {\bf Disappearance channels} &    &   &   & {\bf Muon beam option} \\ \hline
\rule{0pt}{3ex} Signal: & \multirow{2}{*}{$\bar{\nu}_\mu \rightarrow \bar{\nu}_\mu$ and $\nu_e \rightarrow \nu_\mu$} & & Background: & \multirow{2}{*}{$\bar{\nu}_\mu$ NC} \\ 
\rule{0pt}{3ex}         & & & & \\ 
\rule{0pt}{3ex}         & \multirow{2}{*}{$\nu_\mu \rightarrow \nu_\mu$ and $\bar{\nu}_e \rightarrow \bar{\nu}_\mu$} &  &  & \multirow{2}{*}{$\nu_\mu$ NC} \\ 
\rule{0pt}{3ex}         & & & & \\ \hline
\rule{0pt}{3ex} {\bf Appearance channels} &    &   &   & {\bf Beta beam option} \\ \hline
\rule{0pt}{3ex} Signal: & \multirow{2}{*}{$\nu_e \rightarrow \nu_\mu$} & & Background: & \multirow{2}{*}{$\nu_e$ NC} \\ 
\rule{0pt}{3ex}         & & & & \\ 
\rule{0pt}{3ex}         & \multirow{2}{*}{$\bar{\nu}_e \rightarrow \bar{\nu}_\mu$} & & & \multirow{2}{*}{$\bar{\nu}_e$ NC} \\ 
\rule{0pt}{3ex}         & & & & \\ 
\rule{0pt}{3ex}         & \multirow{2}{*}{$\nu_e \rightarrow \nu_\tau$} & & & \multirow{2}{*}{$\nu_e$ NC} \\ 
\rule{0pt}{3ex}         & & & & \\ \hline
\rule{0pt}{3ex} {\bf Disappearance channels} &    &   &   & {\bf Beta beam option} \\ \hline
\rule{0pt}{3ex} Signal: & \multirow{2}{*}{$\nu_e \rightarrow \nu_e$} & & Background: & \multirow{2}{*}{$\nu_e$ NC} \\ 
\rule{0pt}{3ex}         & & & & \\ 
\rule{0pt}{3ex}         & \multirow{2}{*}{$\bar{\nu}_e \rightarrow \bar{\nu}_e$} &  &  & \multirow{2}{*}{$\bar{\nu}_e$ NC} \\ 
\rule{0pt}{3ex}         & & & & \\ \hline
\end{tabular}}
\end{center}
\end{table}

The simulation study is carried out with General Long-Baseline Experiment Simulator~\cite{Huber:2004ka,Huber:2007ji} and its New Physics package~\cite{Kopp:2006wp}. We begin this study by analysing the prospects of all available baseline setups while considering muon energies between 15 and 50~GeV, which have previously been considered in e.g. Refs.~\cite{Tang:2009wp,Tang:2009na,Tang:2017khg}. We test the available beam configurations for $E_\mu =$ 15, 25 and 50~GeV. As for the neutrino detector, we adopt the M.I. and E.C.C. hybrid detector setup assuming 50~kton fiducial mass and 10 years of data taking. The operational time is divided evenly between the positively and negatively charged muon modes. We also study the beta beam option with ion acceleration factors $\gamma =$ 200, 500 and 1000. Neutrino detector of L.Sc. technology and 50~kton fiducial mass are chosen for this option. The key details regarding the simulated configurations are summarized for the muon and beta beam options in table\,\ref{tab:configuration}.

The channel compositions of the considered neutrino beam setups are shown in table\,\ref{tab:channels}. In the muon beam option, signal events in the appearance channels consist of neutrinos undergoing charged current (CC) interactions in the gold channels $\nu_e \rightarrow \nu_\mu$ and $\bar{\nu}_e \rightarrow \bar{\nu}_\mu$ or the silver channel $\nu_e \rightarrow \nu_\tau$. The efficiencies for these channels are 45\%, 35\% and 9.6\%, respectively. Main backgrounds to the gold channels are the neutral currents (NC) and charge mis-identifications (mis-id.), where acceptance rates are taken to be 5$\times$10$^{-6}$ for each. The silver channel on the other hand acquires backgrounds from a number of CC and NC channels, the largest component being CC events from $\bar{\nu}_\mu \rightarrow \bar{\nu}_\tau$ with 0.1\% acceptance. The disappearance channels $\nu_\mu \rightarrow \nu_\mu$ and $\bar{\nu}_\mu \rightarrow \bar{\nu}_\mu$ are analysed while no charge-identification is assumed\footnote{An alternative to the joint analysis is to introduce charge identification. In such case, efficiencies to $\nu_\mu$ and $\bar{\nu}_\mu$ events drop to 45\% and 35\%, respectively.}, increasing the efficiency to 90\% in both channels. Disappearance channels gain backgrounds from $\nu_\mu$ and $\bar{\nu}_\mu$ NC events at 10$^{-5}$ rate. In the beta beam option, signal events consist of $\nu_e \rightarrow \nu_\mu$, $\bar{\nu}_e \rightarrow \bar{\nu}_\mu$ and $\nu_e \rightarrow \nu_\tau$ CC events as well as $\nu_e \rightarrow \nu_e$ and $\bar{\nu}_e \rightarrow \bar{\nu}_e$ CC events. The efficiencies for $\nu_\mu$, $\bar{\nu}_\mu$ and $\nu_\tau$ events in the appearance channels are 80\%, 20\% and 3.56\%, respectively. The efficiencies for $\nu_e$ and $\bar{\nu}_e$ in the disappearance channels are both 20\%. The main backgrounds to the {\color{black}beta} beam option are NC events, which are treated with 0.1\% acceptance rate.

\subsection{Simulation methods}
\label{sec:conf2}

In this subsection, we describe the numerical methods used in the analysis of the simulated neutrino oscillation data. The simulated data is analysed with the following $\chi^2$ function:
\begin{equation}
\chi^2 =  \sum_{i} 2\left[ T_{i} - O_{i} \left( 1 + \log\frac{O_{i}}{T_{i}} \right) \right] + \frac{\zeta_{\text{sg}}^2}{\sigma_{\zeta_{\text{sg}}}^2} + \frac{\zeta_{\text{bg}}^2}{\sigma_{\zeta_{\text{bg}}}^2} + {\rm priors},
\label{eq:chi2}
\end{equation}
where $O_{i}$ and $T_{i}$ are the number of simulated events for the reference and test data in $i^{\rm th}$ energy bin, computed from the true and test values respectively. The systematic uncertainties are taken into account with the pull method~\cite{Fogli:2002pt} by using the pull parameters $\zeta_{\text{sg}}$ and $\zeta_{\text{bg}}$, {\color{black}which describe the systematic uncertainties in the signal and background events. The systematic uncertainties are uncorrelated between the oscillation channels.} In the muon beam option, the systematic uncertainty treatment is as follows: Muon appearance channels $\nu_e \rightarrow \nu_\mu$ and $\bar{\nu}_e \rightarrow \bar{\nu}_\mu$ are characterised with 2.5\% normalization uncertainty on the signal events, whereas 15\% uncertainty is assumed for the tau neutrino appearance signal $\nu_e \rightarrow \nu_\tau$. An overall 20\% uncertainty is adopted for the analysis of background events. In the analysis of the beta beam facility, a uniform 2.5\% normalization uncertainty is imposed on the signal events and 5\% on the background events. Tau neutrino appearance is treated with 2.5\% signal and 20\% background uncertainties in the beta beam option.

In the analysis of the simulated data, we make use of the neutrino oscillation data that has been obtained in previous experiments. We adopt priors from the global three-neutrino best-fit values presented by the NuFit group~\cite{NuFIT:5-0}. The approximate values for the central values and relative errors at 1\,$\sigma$ and 3\,$\sigma$ confidence levels (CL) are shown in table\,\ref{tab:bestfits}. In our analysis, we adopt the prior values as Gaussian distributions defined by the central values and 1\,$\sigma$ CL errors as given in the table. Without loss of generality, we carry out our analysis assuming normally ordered neutrino masses, that is, $m_3 \gg m_2 > m_1$.

\begin{table}[!t]
\caption{\label{tab:bestfits} The best-fit values of the standard oscillation parameters presented with 1$\,\sigma$ and 3$\,\sigma$ CL relative errors~\cite{NuFIT:5-0,Esteban:2020cvm}. The values are shown for normal mass ordering.}
\begin{center}
\begin{tabular}{|c|c|c|c|c|}\hline
{\bf Parameter} & {\bf Central value} & {\bf Relative error (1\,$\sigma$)} & {\bf Relative error (3\,$\sigma$)} \\ \hline
\rule{0pt}{3ex}$\theta_{12}$ ($^\circ$) & 33.4 & 2.3\% & 13.7\% \\ \hline
\rule{0pt}{3ex}$\theta_{13}$ ($^\circ$) & 8.6 & 1.4\% & 8.5\% \\ \hline
\rule{0pt}{3ex}$\theta_{23}$ ($^\circ$) & 49.2 & 2.1\% & 25.3\% \\ \hline
\rule{0pt}{3ex}$\delta_\text{CP}$ ($^\circ$) & 197.0 & 12.9\% & unconstrained \\ \hline
\rule{0pt}{3ex}$\Delta m_{21}^2$ (10$^{-5}$ eV$^2$) & 7.4 & 2.8\% & 16.4\% \\ \hline
\rule{0pt}{3ex}$\Delta m_{31}^2$ (10$^{-3}$ eV$^2$) & 2.5 & 1.1\% & 6.5\% \\ \hline
\end{tabular}
\end{center}
\end{table}

The simulated data is analysed in the {\em CP} violation search, precision measurements of $\theta_{23}$, $\delta_{CP}$ and $\Delta m_{31}^2$ as well as sensitivities to non-unitarity parameters $\alpha_{ij}$ and NSI parameters $\epsilon_{\ell \ell'}^m$. Standard oscillation probabilities are calculated numerically with GLoBES, while the non-unitarity and NSI effects are evaluated with self-developed probability code and New Physics package, respectively. It is also useful to compare the expected prospects in PROMPT and other simulated configurations with those of T2HK and DUNE. For this reason, we simulate DUNE and T2HK by following configurations and techniques described in Refs.~\cite{DUNE:2016ymp,Du:2021rdg}.

\section{Physics prospects}
\label{sec:pheno}

In this section, we present the numerical result of this work. We begin this study in section\,\ref{sec:pheno:muon_beam} by calculating the potential to discover {\em CP} violation and measure the standard oscillation parameters $\theta_{23}$, $\delta_{CP}$ and $\Delta m_{31}^2$. The muon and beta beam options are simulated with various baseline lengths and neutrino beam properties. We also examine the prospects to study non-unitarity in neutrino mixing and non-standard neutrino interactions. The study is continued in section\,\ref{sec:pheno:PROMPT} where projections are presented for the PROMPT setup.

\subsection{Optimization of the neutrino beam and baseline length}
\label{sec:pheno:muon_beam}

The accelerator and detector laboratories discussed in this work give access to a variety of baseline lengths between 162~km and 1871~km. In order to determine the most suitable beam properties, we simulate the muon beam option with three different muon beam energies $E_\mu =$ 15\,GeV, 25\,GeV and 50\,GeV. The beta beam option is correspondingly studied with three different ion acceleration factors $\gamma =$ 200, 500 and 1000. 

\begin{figure}[!t]
        \center{\includegraphics[width=\textwidth]{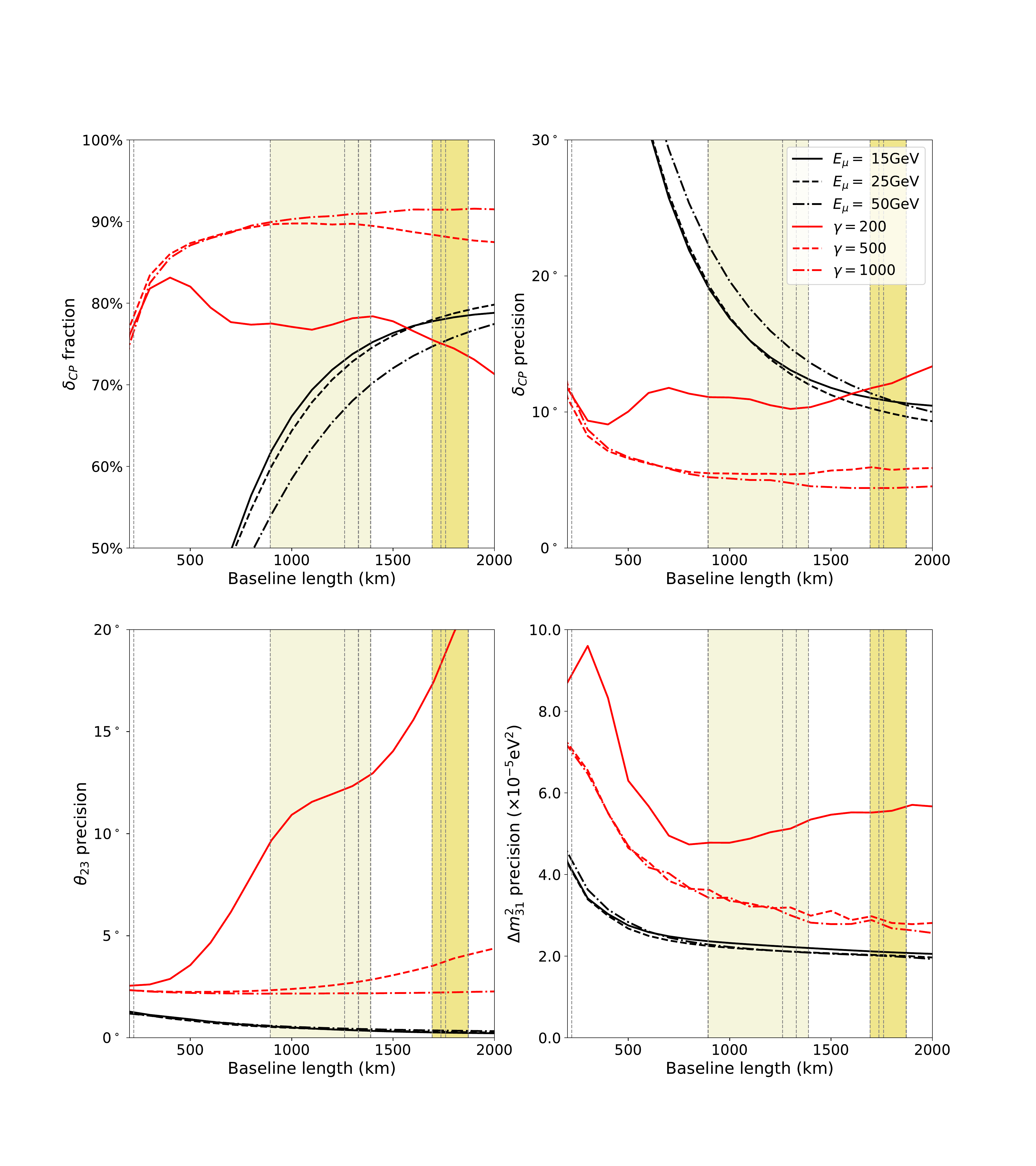}}
        \caption{Sensitivities to {\em CP} violation and precision on the mixing parameters $\theta_{23}$, $\Delta m_{31}^2$ and $\delta_{CP}$ in the considered neutrino beam setup with baseline lengths 100~km...2000~km. Projections are based on experiment setups driven by muon beams (black curves) and beta beams (red curves). Muon beam energies $E_\mu = $15, 25, 50~GeV and gamma factors $\gamma =$ 200, 500, 1000 are shown.}
        \label{fig:prec-opt}
\end{figure}

In figure\,\ref{fig:prec-opt}, the expected sensitivities to {\em CP} violation (top left), $\delta_{CP}$ precision (top right), $\theta_{23}$ precision (bottom left) and $\Delta m_{31}^2$ precision (bottom right) are shown as function of baseline length within [200,~2000]~km range. In the {\em CP} violation panel, the discovery potential is expressed as a fraction of theoretically allowed $\delta_{CP}$ values for which discovery could be reached by at least 3$\,\sigma$ CL. In the $\delta_{CP}$, $\theta_{23}$ and $\Delta m_{31}^2$ precision panels, {\color{black}the achievable uncertainty for each parameter is presented at 1$\,\sigma$ CL.} The expected sensitivities to each observable is presented with black curves for muon beam option and with red curves for the beta beam option, respectively. The baseline length that correspond to the various accelerator-detector laboratory pairs represented by the vertical dashed lines. The laboratory pairs can be divided into two groups, which are indicated by the shaded rectangles. The medium-baseline group at the center of the parameter space covers the laboratory pairs with baseline lengths 894~km...1389~km. This region contains four configurations: CAS-IMP$\rightarrow$CJPL, Nanjing$\rightarrow$JUNO CSNS$\rightarrow$CJPL and CiADS$\rightarrow$CJPL. The long-baseline group on the other hand consists of laboratory pairs with baseline lengths 1693~km...1814~km. The long-baseline region includes configurations Nanjing$\rightarrow$CJPL, SPPC$\rightarrow$CJPL, CAS-IMP$\rightarrow$JUNO and SPPC$\rightarrow$JUNO respectively. 

The sensitivities shown in figure\,\ref{fig:prec-opt} suggest that the beta beam option is generally more sensitive to {\em CP} violation and the relative precision on $\delta_{CP}$ than the muon beam option. The muon beam option on the other hand performs better with the precision measurements of $\theta_{23}$ and $\Delta m_{31}^2$. Regarding the considered baseline lengths, the interesting region is where the turning point occurs. After the turning point, any increment in baseline length leads to insignificant improvements or even decrease in the expected sensitivity. For the muon beam option, the turning point is located within the long-baseline group in the {\em CP} violation panel and within the medium-baseline group in the precision measurement panels. For the beta beam option, the turning point occurs at baseline lengths much shorter than the medium-baseline and long-baseline groups. We found no notable contribution from the tau neutrino appearance channel in the sensitivities. If one considers baseline lengths in the medium-baseline group, a muon beam of either 15~GeV or 25~GeV and a beta beam of $\gamma =$ 500 or 1000 yield the best sensitivities.

\begin{figure}[!t]
        \center{\includegraphics[width=\textwidth]{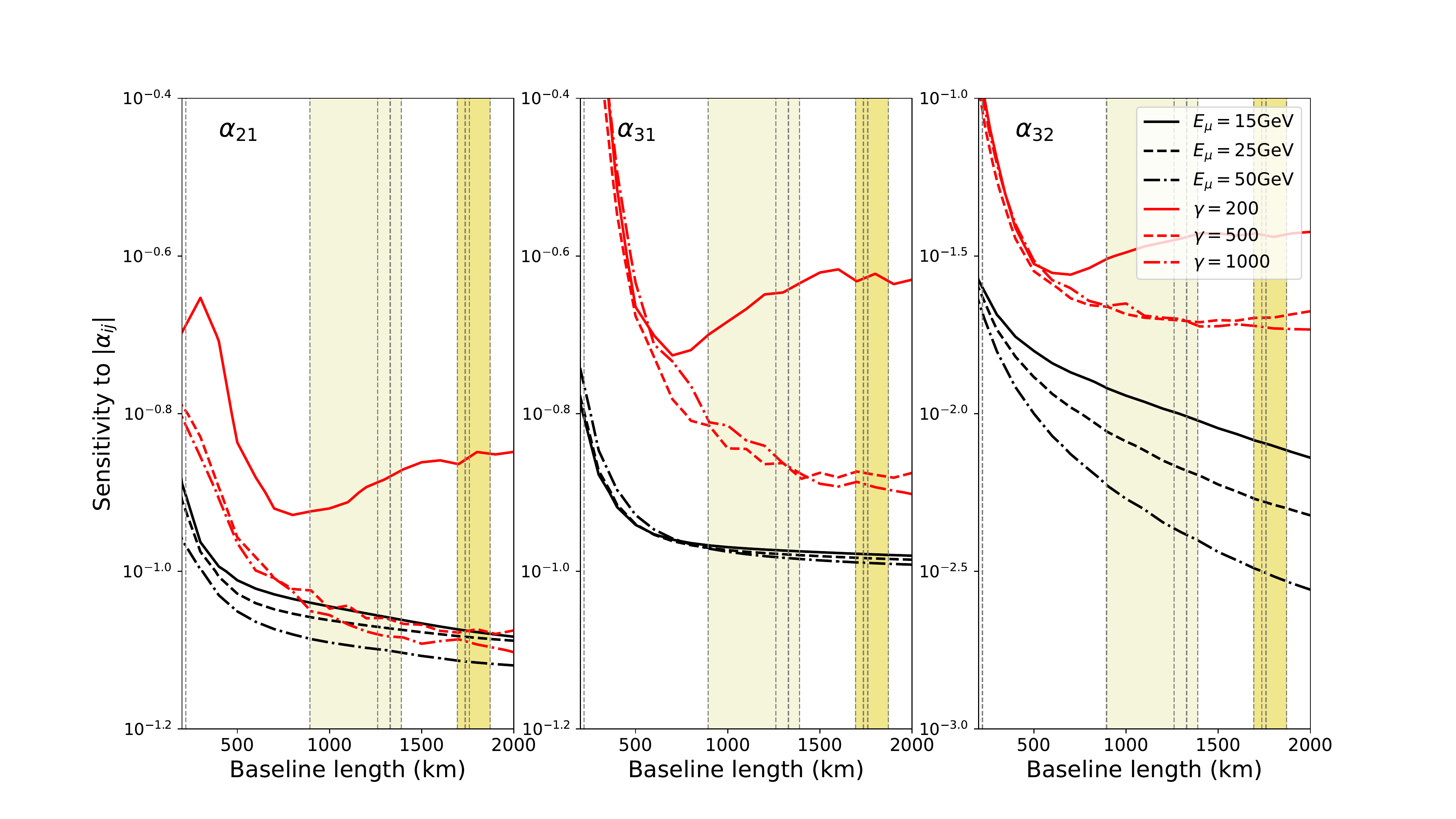}}
        \caption{Sensitivities to the non-unitarity parameters $\alpha_{i j}$ ($i$, $j = $ 1, 2, 3) in the considered neutrino beam setups with baselines between 100~km and 2000~km. Configurations based on muon beam energies $E_\mu =$ 5, 25 and 50~GeV are shown with black curves and ion acceleration factors $\gamma =$ 200, 500 and 1000 are shown with red curves, respectively.}
        \label{fig:alpha-opt}
\end{figure}

The enhanced precision on the standard oscillation parameters also gives an opportunity to test unitarity of the PMNS mixing matrix. In this work, we parameterize deviations from unitarity paradigm with parameters $\alpha_{i j}$ ($i$, $j =$ 1, 2, 3) introduced in section\,\ref{sec:prob2}. In the experiment configurations considered in this work, the constraints on the magnitudes of off-diagonal parameters $\alpha_{21}$, $\alpha_{31}$ and $\alpha_{32}$ are presented in figure\,\ref{fig:alpha-opt}. In each panel, the sensitivity to each parameter is shown while keeping other non-unitarity parameters fixed at the value that corresponds to the unitarity paradigm. The results are shown at 90\% CL. The sensitivities obtained in this method show good prospects to study non-unitarity in the medium-baseline and long-baseline groups. In the muon beam option, the turning point occurs at baseline lengths around 500~km for parameters $\alpha_{21}$ and $\alpha_{31}$, whereas for parameter $\alpha_{32}$ the turning point is not visible in the considered parameter space. In the beta beam option, the turning point is achieved about 700~km...800~km for $\gamma =$ 200 and 1000~km...1200~km, for $\gamma =$ 500 and 1000. The highest sensitivities to study non-unitarity parameters are found with the muon beam option, where laboratory pairs in the long-baseline group appear to yield the most stringent constraints. Regarding the effect of tau neutrino appearance, we found modest improvements in the sensitivities to $\alpha_{31}$ and $\alpha_{32}$ in both beam options.

\begin{figure}[!t]
        \center{\includegraphics[width=\textwidth]{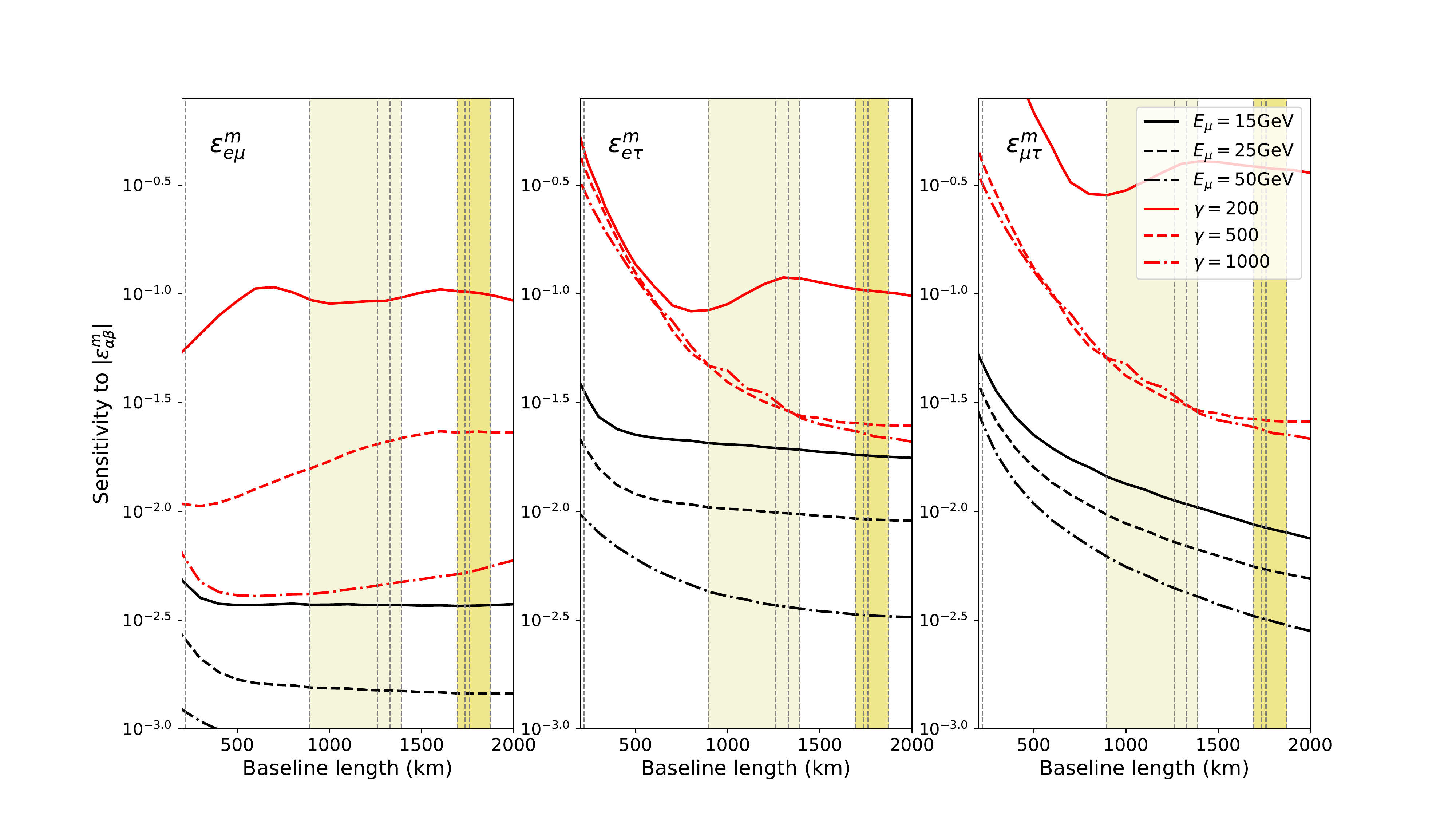}}
        \caption{Sensitivities to the non-standard interaction parameters $\epsilon^m_{\ell \ell'}$, where $\ell$, $\ell' = e$, $\mu$, $\tau$ in the neutrino beam setup with baselines between 100~km and 2000~km. Muon beam setups of 15~GeV, 25~GeV and 50~GeV parent energies are shown with black curves and beta beam setups of 200, 500 and 1000 acceleration factors with red curves, respectively.}
        \label{fig:epsilon-opt}
\end{figure}

The final physics topic to be discussed in this work is the search for non-standard interactions in the neutrino sector. In the considered experiment configurations, the most relevant form of neutrino NSI is formed in the propagation. For this purpose, we leave out potential neutrino NSI attributed to the neutrino production and detection processes and focus on the constraining ability of the matter NSI parameters $\epsilon_{\ell \ell'}^m = |\epsilon_{\ell \ell'}^m| e^{-i \phi_{\ell \ell'}^m}$ where $\ell$, $\ell' = e$, $\mu$ and $\tau$. The sensitivities to the off-diagonal matter NSI parameters $\epsilon_{e \mu}^m$, $\epsilon_{e \tau}^m$ and $\epsilon_{\mu \tau}^m$ are calculated assuming only one matter NSI parameter to be non-zero at a time. Focusing only on the magnitudes of the matter NSI parameters, we take each parameter to be real. The sensitivities to $|\epsilon_{e \mu}^m|$, $|\epsilon_{e \tau}^m|$ and $|\epsilon_{\mu \tau}^m|$ are shown in figure\,\ref{fig:epsilon-opt}. The turning points in the muon beam configurations occur at baseline lengths of about 300~km...500~km for $|\epsilon_{e \mu}^m|$ and $|\epsilon_{e \tau}^m|$ and at baseline lengths much longer than 2000~km for $|\epsilon_{\mu \tau}^m|$. This makes the long-baseline group more favourable for studying the matter NSI parameters. In the beta beam option, the medium-baseline group is favoured over the long-baseline group for acceleration factor $\gamma =$ 200, whereas the long-baseline group is preferred for $\gamma =$ 500 and 1000. The most stringent constraints are achieved with high muon beam energies and high ion acceleration factors.

We also investigated the effect of the tau neutrino appearance channel $\nu_e \rightarrow \nu_\tau$ in the sensitivities to the off-diagonal matter NSI parameters. We found significant improvement in the sensitivity to $\alpha_{31}$ and modest improvement to the sensitivity to $\alpha_{32}$ in the muon beam option. In the beta beam option, we found little or no effect arising from the $\nu_e \rightarrow \nu_\tau$ channel.

We finally comment on the accelerator-detector laboratory pairs investigated in figures~\ref{fig:prec-opt}--\ref{fig:epsilon-opt}. Basing on the obtained results, we find the setups both in the medium-baseline and long-baseline groups able to offer competitive choices for a future accelerator-based neutrino beam experiment in China. In the muon beam option, we find the baseline lengths enclosed in the long-baseline group more sensitive to {\em CP} violation, non-unitarity parameter $\alpha_{32}$ and matter NSI parameter $\epsilon_{\mu \tau}^m$, whilst the differences between the two baseline groups are similar for the other parameters. The sensitivity to non-unitarity and matter NSI parameters increase linearly along with the muon beam energy. In the beta beam option, we find a mild preference for the medium-baseline group for ion acceleration factors $\gamma =$ 200 and for the long-baseline group for $\gamma =$ 500 and 1000. We therefore find the experiment configurations represented in the long-baseline group suitable for a balanced physics program. {\color{black}We also analysed sensitivities in alternative setups where muon beam energies $E_\mu =$ 20 and 100~GeV as well as ion acceleration factors $\gamma =$ 300 and 400 were considered, finding consistent results. Higher values of $E_\mu$ and $\gamma$ generally lead to higher statistics near the relevant oscillation maxima, meaning also higher sensitivities to the oscillation parameters. As one can see from the figures, statistics plays an important role in the muon beam setup for the majority of parameters, while a careful selection of muon beam energy is necessary to achieve the largest fraction of $\delta_{CP}$ in {\em CP} violation discovery. The beta beam option on the other hand benefits from higher ion acceleration factors up to about $\gamma =$ 500 with all parameters except $\epsilon_{e \mu}^m$, where $\gamma =$ 1000 yields a significantly higher sensitivity.}

\subsection{Sensitivities in PROMPT}
\label{sec:pheno:PROMPT}

We now present a case study on PROMPT, where a high-power neutrino beam facility based on decay-in-flight muons is proposed to the SPPC injector chain and hybrid detector based on magnetized iron and emulsion cloud chamber techniques to the CJPL site. Based on the results obtained in the previous subsection, we simulate PROMPT with the {\color{black}muon} beam setup of 25~GeV parent energy and with the SPPC$\rightarrow$CJPL baseline.

\subsubsection*{Precision measurements on standard oscillation parameters}
\label{sec:pheno:CPV}

  \begin{figure}[!t]
        \center{\includegraphics[width=1.\textwidth]
        {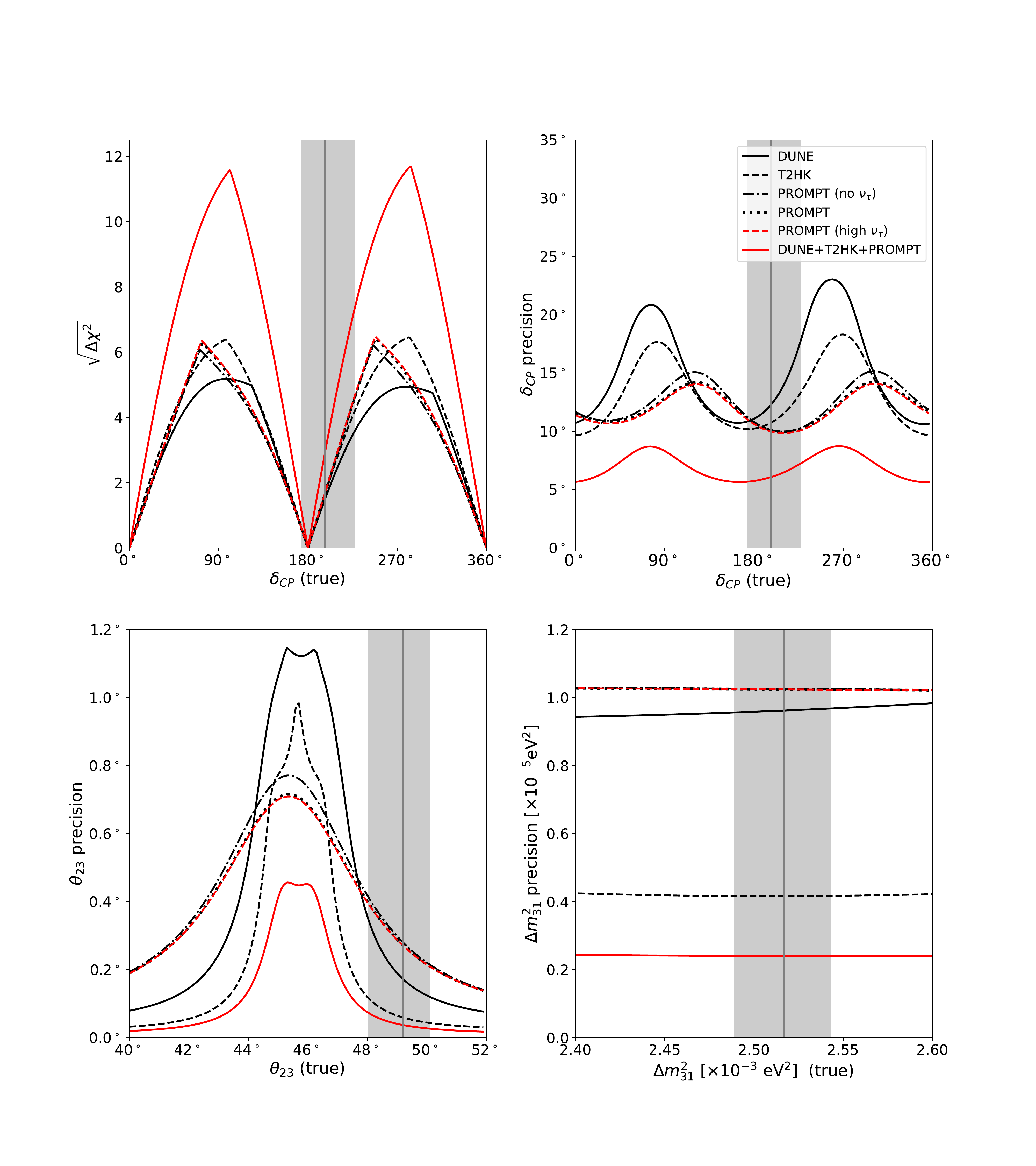}}
        \caption{\label{fig:prec_prompt} Expected sensitivities to {\em CP} violation (top left) and precision on $\delta_{CP}$ (top right), $\theta_{23}$ (bottom left) and $\Delta m_{31}^2$ (bottom right). Also shown are the expected sensitivities for the configuration without $\nu_\tau$ appearance {\color{black}(no $\nu_\tau$) and with elevated $\nu_\tau$ efficiency (high $\nu_\tau$)} as well as for DUNE and T2HK setups. The sensitivities are given at 1$\,\sigma$ CL while the global fit result with 1$\,\sigma$ CL uncertainties are indicated by the shaded regions.}
      \end{figure}
      
We begin by simulating PROMPT in the standard three-neutrino oscillation picture. Discovery potential to {\em CP} violation and expected precision on parameters $\delta_{CP}$, $\theta_{23}$ and $\Delta m_{31}^2$ are presented for the 25~GeV muon beam setup in figure\,\ref{fig:prec_prompt}. The top left panel shows the {\em CP} violation discovery potential within theoretically allowed $\delta_{CP}$ values, whereas other panels present {\color{black}the absolute precision} on parameters $\delta_{CP}$ (top right), $\theta_{23}$ (bottom left) and $\Delta m_{31}^2$ (bottom right). In each panel, sensitivities are shown for DUNE, T2HK and PROMPT configurations as well as for their combination.  {\color{black}We also display the discovery potentials in the case where the sensitivity to $\nu_\tau$ in PROMPT is either turned off (no $\nu_\tau$) and when it is significantly increased  (high $\nu_\tau$). For the latter, we adopt a $\nu_\tau$ efficiency that differs from the baseline setup by a factor of 10}. Finally, we depict the experimentally allowed values at 1$\,\sigma$ CL from the recent global fit~\cite{Esteban:2020cvm}. The currently allowed values are indicated with the shaded regions, whereas the vertical grey lines represent the corresponding best-fit values.

As one can see from the obtained results, the sensitivities obtained for the PROMPT setup are comparable with the sensitivities projected for the future long-baseline neutrino experiments T2HK and DUNE. Figure\,\ref{fig:prec_prompt} reveals that {\em CP} violation can be discovered in T2HK at 6.2$\,\sigma$ CL when the true value  of $\delta_{CP}$ is 282$^\circ$. Similar discovery reach can be achieved in PROMPT when $\delta_{CP} \simeq$ 248$^\circ$. The lowest sensitivity is obtained in DUNE, where {\em CP} violation can be discovered at about 5.2$\,\sigma$ CL when $\delta_{CP} \simeq$ 97$^\circ$. In the precision measurement of $\delta_{CP}$, the most stringent constraints are achieved in the PROMPT setup, whereas the next-stringent constraints are obtained for the T2HK and DUNE configurations, respectively. {\color{black}In the PROMPT setup, $\delta_{CP}$ can be measured by as low as 14.2$^\circ$ uncertainty at 1$\,\sigma$ CL, whereas the corresponding number is 18.3$^\circ$ for T2HK and 23.0$^\circ$ for DUNE. In case of $\theta_{23}$, 0.7$^\circ$, 1.0$^\circ$ and 1.1$^\circ$ uncertainties are obtained respectively for PROMPT, T2HK and DUNE. For $\Delta m_{31}^2$, the precisions are 1.0, 0.4 and 0.9$\times$10$^{-5}$eV$^2$.} The effect of $\nu_\tau$ sample in PROMPT is relatively small in all four panels of figure\,\ref{fig:prec_prompt}.

The true potential of PROMPT becomes evident when the PROMPT setup is simulated together with the T2HK and DUNE configurations. In figure\,\ref{fig:prec_prompt}, the combined sensitivities are indicated with the solid red curves. In the precision measurement of $\delta_{CP}$, the combined run of PROMPT, T2HK and DUNE results in {\color{black}8.7$^\circ$ precision at 1$\,\sigma$ CL}.

In summary, we find the sensitivities obtained with the PROMPT setup very promising for the measurement of the standard oscillation parameters $\delta_{CP}$, $\theta_{23}$ and $\Delta m_{31}^2$. The most significant contributions to the sensitivities of the PROMPT setup are found in the golden channels $\nu_e \rightarrow \nu_\mu$ and $\bar{\nu}_e \rightarrow \bar{\nu}_\mu$. The silver channel $\nu_e \rightarrow \nu_\tau$ is also found to have a noticeable effect, although its impact is relatively small.

\subsubsection*{Non-unitarity of the neutrino mixing matrix}
\label{sec:pheno:NU}

\begin{figure}[!t]
        \center{\includegraphics[width=\textwidth]{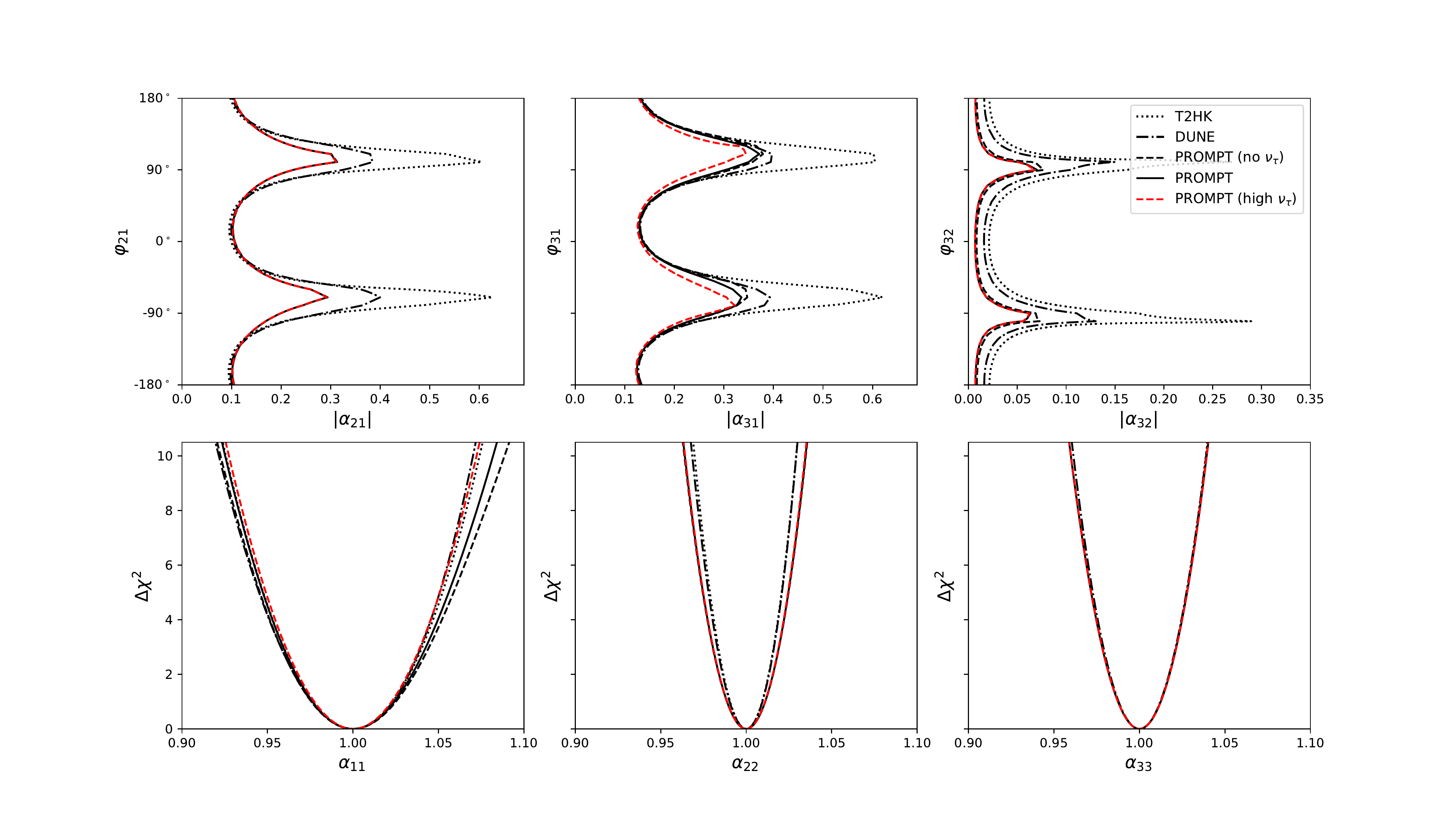}}
        \caption{The sensitivity to the non-unitarity parameters $\alpha_{i j} = |\alpha_{i j}| e^{-i \varphi_{i j}}$ ($i$, $j =$ 1, 2 and 3) in PROMPT. The projections of PROMPT are presented {\color{black} for the baseline setup (solid), without sensitivity to tau neutrinos (dashed) and with elevated $\nu_\tau$ efficiency (colour dashed), respectively.} The expected sensitivities for the T2HK (dotted) and DUNE (dot-dashed) are also shown. Sensitivities to off-diagonal parameters are presented at 90\% CL.}
        \label{fig:alpha}
\end{figure}

We now simulate the PROMPT setup {\color{black}in case of} non-unitary mixing.  We consider both the off-diagonal parameters $\alpha_{21}$, $\alpha_{31}$ and $\alpha_{32}$ and the diagonal parameters $\alpha_{11}$, $\alpha_{22}$ and $\alpha_{33}$. The exclusion limits these parameters are presented for the PROMPT setup in figure\,\ref{fig:alpha}. The top row of the panels represents the sensitivities to the off-diagonal parameters, while the sensitivities to the diagonal parameters are shown in the bottom row. The sensitivities are provided for the PROMPT setup both with and without the sensitivity to tau neutrino appearance. The corresponding sensitivities are also shown for the T2HK and DUNE configurations. 

As one can see from figure\,\ref{fig:alpha}, the most sensitive probes to the off-diagonal parameters are provided respectively by the PROMPT, DUNE and T2HK configurations. In case of $\alpha_{21}$, the T2HK, DUNE and PROMPT setups result in 0.62, 0.40 and 0.31 upper limits, respectively. In case of $\alpha_{31}$, the corresponding sensitivities are 0.62, 0.40 and 0.31, whereas in case of $\alpha_{32}$ the sensitivities are 0.29, 0.15 and 0.07. The contribution from the tau neutrino appearance channel in the PROMPT setup results in small improvements in the sensitivities to $\alpha_{31}$ and $\alpha_{32}$, where the expected $\nu_\tau$ sample tightens the constraints on $|\alpha_{31}|$ and $|\alpha_{32}|$ near the maximally {\em CP}--violating values $\varphi_{31}$, $\varphi_{32} \simeq \pm$90$^\circ$. The sensitivities to the diagonal parameters show only little difference. For the diagonal parameters, notable differences between the sensitivities are observed only for $\alpha_{11}$.

{\color{black}The relatively low contribution from the tau neutrino appearance channel $\nu_e \rightarrow \nu_\tau$ can be traced back to the detector efficiency to tau neutrinos. The present challenges in the reconstruction of $\tau^-$ set limit to the $\nu_\tau$ statistics that can be acquired in the neutrino detector. Owing to the lower efficiency, the sensitivities that can be obtained from the $\nu_e \rightarrow \nu_\tau$ channel are comparatively lower than the ones corresponding to $\nu_e \rightarrow \nu_\mu$.} {\color{black}For comparison, sensitivities for the PROMPT setup are also shown without $\nu_\tau$ sensitivity as well as with enhanced $\nu_\tau$ sensitivity.}

We find PROMPT to have altogether a very good ability to probe the non-unitarity parameters. The expected constraints on the three off-diagonal parameters are notably stricter than those predicted for the T2HK and DUNE configurations.

\subsubsection*{Non-standard neutrino interactions}
\label{sec:pheno:NSI}

We finally discuss the sensitivities to the matter NSI parameters in PROMPT. As before, we ignore the effects from the source and detection NSI and focus only on the matter NSI parameters\footnote{It is safe to ignore source and detection NSI effects in long-baseline neutrino experiments up to the point where sensitivity to matter NSI becomes comparable with source and detection NSI parameters, which are currently constrained to percent level~\cite{Biggio:2009nt}. Beyond this point, correlations with source and detection NSI parameters must be taken into account to attain a more realistic estimate on sensitivities, as was done in e.g. Ref.~\cite{Blennow:2016etl}.}. We furthermore consider only one matter NSI parameter at a time.

\begin{figure}[!t]
        \center{\includegraphics[width=\textwidth]{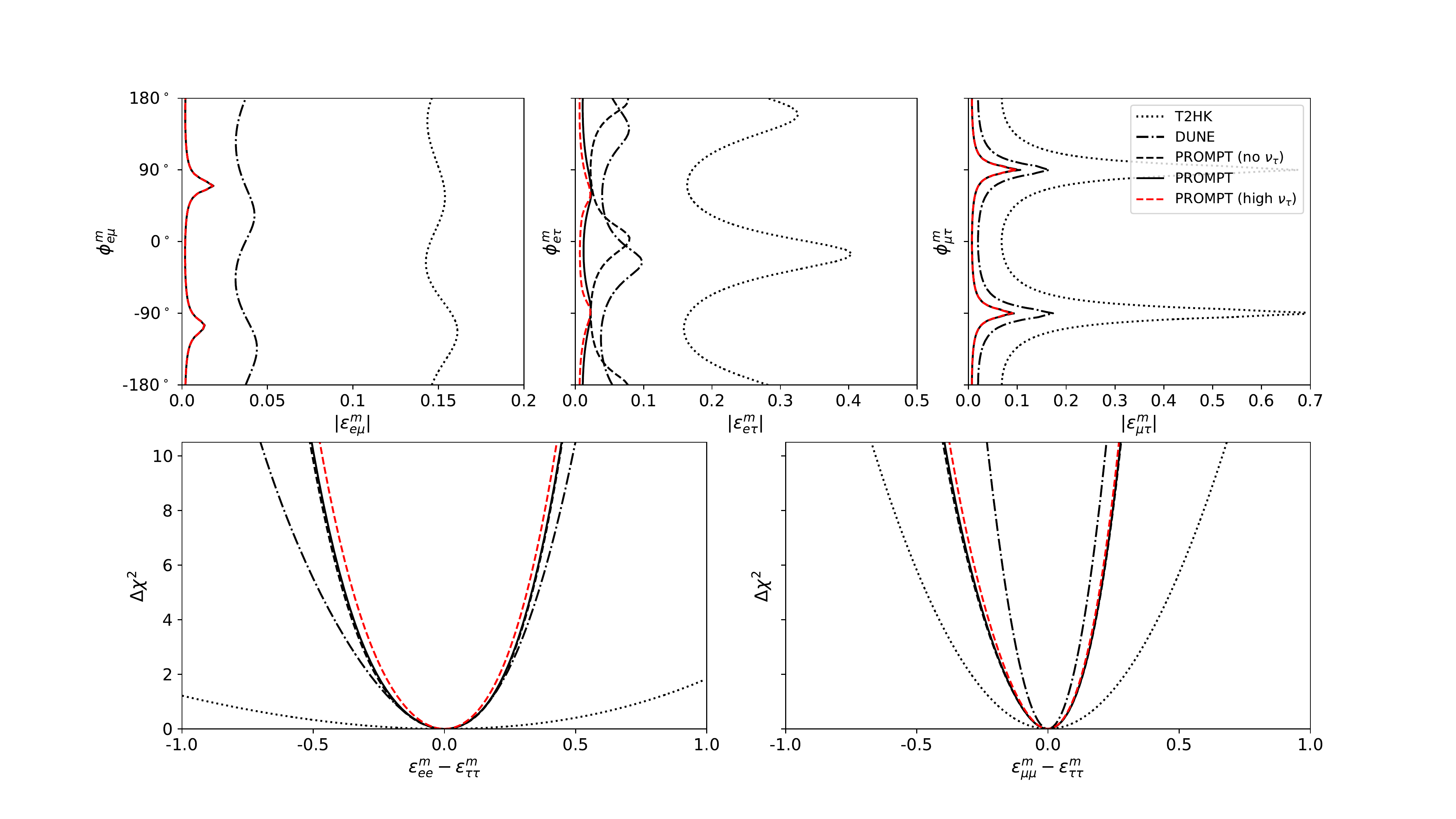}}
        \caption{The sensitivities to the matter NSI parameters $\epsilon^m_{\alpha \beta}$ ($\alpha$, $\beta = e$, $\mu$, $\tau$) as function of the magnitude and phase. The sensitivities to the diagonal elements are shown for ($\epsilon^m_{ee} - \epsilon^m_{\tau\tau}$) and ($\epsilon^m_{\mu\mu} - \epsilon^m_{\tau\tau}$). The projections are presented for PROMPT with the baseline setup (solid), without sensitivity to tau neutrinos (dashed), {\color{black}and with enhanced $\nu_\tau$ sensitivity (colour dashed). Sensitivities are also shown for DUNE (dot-dashed) and T2HK (dotted)}. All sensitivities are presented at 90\% CL.}
        \label{fig:epsilon}
\end{figure}

The sensitivities to the NSI parameters are presented in figure\,\ref{fig:epsilon}. The top three panels represent the exclusion limits for the off-diagonal parameters $\epsilon_{e \mu}$, $\epsilon_{e \tau}$ and $\epsilon_{\mu \tau}$. All sensitivities are projected at 90\% CL. In case of $\epsilon_{e \mu}^m$, PROMPT has clear advantage over T2HK and DUNE setups. As is indicated in the top left panel, the parameter space of $|\epsilon_{e \mu}|$ is constrained below 0.02 in the PROMPT setup. The corresponding sensitivities for the DUNE and T2HK setups are 0.04 and 0.16, respectively. The tau neutrino appearance makes no visible contribution, and the sensitivity of the configuration where tau neutrinos are neglected in PROMPT overlaps with the baseline configuration. In case of $\epsilon_{e \tau}^m$, differences between PROMPT, DUNE and T2HK become more apparent with respective sensitivities 0.03, 0.10 and 0.40. The $\nu_\tau$ appearance in PROMPT has a significant effect in this case: whereas the sensitivity in the baseline configuration is about $|\epsilon_{e \tau}^m| \lesssim$ 0.03, the sensitivity in the configuration without $\nu_\tau$ is 0.10 at 90\% CL\footnote{{\color{black}It is interesting to notice that the sensitivity to $\epsilon_{e \tau}^m$ in matter NSI case is higher than it is to parameter $\alpha_{31}$ in the non-unitary mixing case. The difference between the sensitivities can be explained at least partially by the analytical probabilities for $\nu_e \rightarrow \nu_\tau$ discussed in sections\,\ref{sec:prob2} and \ref{sec:prob3}.}}. Regarding the final off-diagonal parameter $\epsilon_{\mu \tau}^m$, the tau neutrino appearance channel has no effect in the sensitivity to PROMPT. The expected sensitivities in the PROMPT, DUNE and T2HK setups to $|\epsilon_{\mu \tau}|$ are 0.11, 0.17 and 0.53, respectively.

We study the diagonal NSI parameters $\epsilon_{e e}^m$, $\epsilon_{\mu \mu}^m$ and $\epsilon_{\tau \tau}^m$ through the quantities $\epsilon_{e e}^m - \epsilon_{\tau \tau}^m$ and $\epsilon_{\mu \mu}^m - \epsilon_{\tau \tau}^m$. The sensitivities to $\epsilon_{e e}^m - \epsilon_{\tau \tau}^m$ and $\epsilon_{\mu \mu}^m - \epsilon_{\tau \tau}^m$ in the T2HK, DUNE and PROMPT setups are shown in the bottom row of figure\,\ref{fig:epsilon}. At 90\% CL, the T2HK setup can be expected to yield the least constraining sensitivity, -2.21 $\lesssim \epsilon_{e e}^m - \epsilon_{\tau \tau}^m \lesssim$ 2.29. In the PROMPT and DUNE setups, the corresponding sensitivities are -0.29 $\lesssim \epsilon_{e e}^m - \epsilon_{\tau \tau}^m \lesssim$ 0.26 and -0.34 $\lesssim \epsilon_{e e}^m - \epsilon_{\tau \tau}^m \lesssim$ 0.27, respectively. On the other hand, the constraints on $\epsilon_{\mu \mu}^m - \epsilon_{\tau \tau}^m$ are found to be [-0.11,~0.12], [-0.19,~0.16] and [-0.34,~0.27] in DUNE, PROMPT and T2HK, respectively. The tau neutrino appearance in PROMPT bears no significant effect on the sensitivities to $\epsilon_{e e}^m - \epsilon_{\tau \tau}^m$ or $\epsilon_{\mu \mu}^m - \epsilon_{\tau \tau}^m$.

\section{Summary}
\label{sec:summ}
\begin{table}[!t]
\caption{\label{tab:nsi_prosp} Summary of the expected sensitivities obtained for the PROMPT, T2HK and DUNE setups in this work. The results are shown for the relative precisions on $\delta_{CP}$, $\theta_{23}$ and $\Delta m_{31}^2$, and the allowed values of non-unitarity parameters $\alpha_{i j}$ ($i$, $j =$ 1, 2, 3) and matter NSI parameters $\epsilon^m_{\ell \ell'}$ ($\ell$, $\ell' = e$, $\mu$, $\tau$). {\color{black}The sensitivities are provided at 1\,$\sigma$ CL for $\delta_{CP}$, $\theta_{23}$ and $\Delta m_{31}^2$ and at 90\,\% CL for $|\alpha_{i j}|$ and $|\epsilon^m_{\ell \ell'}|$,} assuming normally ordered neutrino masses. Beam powers are expressed as annual yield of protons on target (POT) for pion-decay-based beams and useful muon decays for muon-decay-based beams.}
\begin{center}
\begin{tabular}{|c|c|c|c|}
\hline
\rule{0pt}{3ex}{\bf Parameter} & {\bf DUNE} & {\bf T2HK} & {\bf PROMPT}  \\ \hline
\rule{0pt}{3ex}$\theta_{23}$ precision {\color{black}[$^\circ$]} & {\color{black}23.0} & {\color{black}18.3} & {\color{black}14.2} \\ \hline
\rule{0pt}{3ex}$\delta_{CP}$ precision {\color{black}[$^\circ$]} & {\color{black}1.1} & {\color{black}1.0} & {\color{black}0.7} \\ \hline
\rule{0pt}{3ex}$\Delta m_{31}^2$ precision {\color{black}[$\times$10$^{-5}$eV$^2$]} & {\color{black}0.9} & {\color{black}0.4} & {\color{black}1.0} \\ \hline
\rule{0pt}{3ex}$\alpha_{11}$ & [0.96,~1.04] & [0.96,~1.04] & [0.96,~1.04] \\ \hline
\rule{0pt}{3ex}$\alpha_{22}$ & [0.98,~1.02] & [0.98,~1.02] & [0.98,~1.02] \\ \hline
\rule{0pt}{3ex}$\alpha_{33}$ & [0.98,~1.02] & [0.98,~1.02] & [0.98,~1.02] \\ \hline
\rule{0pt}{3ex}$|\alpha_{21}|$ & [0.00,~0.40] & [0.00,~0.62] & [0.00,~0.31] \\ \hline
\rule{0pt}{3ex}$|\alpha_{31}|$ & [0.00,~0.40] & [0.00,~0.62] & [0.00,~0.37] \\ \hline
\rule{0pt}{3ex}$|\alpha_{32}|$ & [0.00,~0.15] & [0.00,~0.029] & [0.00,~0.07] \\ \hline
\rule{0pt}{3ex}$\epsilon^m_{e e} - \epsilon^m_{\tau \tau}$ & [-0.34,~0.27] & [-2.21,~2.29] & [-0.29,~0.26] \\ \hline
\rule{0pt}{3ex}$\epsilon^m_{\mu \mu} - \epsilon^m_{\tau \tau}$ & [-0.11,~0.12] & [-0.35,~0.35] & [-0.19,~0.16] \\ \hline
\rule{0pt}{3ex}$|\epsilon^m_{e \mu}|$ & [0,~0.04] & [0,~0.16] & [0,~0.02] \\ \hline
\rule{0pt}{3ex}$|\epsilon^m_{e \tau}|$ & [0,~0.10] & [0,~0.40] & [0,~0.03] \\ \hline
\rule{0pt}{3ex}$|\epsilon^m_{\mu \tau}|$ & [0,~0.17] & [0,~0.053] & [0,~0.11] \\ \hline
\rule{0pt}{3ex}Production method & pion decay & pion decay & muon decay \\ \hline
\rule{0pt}{3ex}Beam power [year$^{-1}$] & 8.82$\times$10$^{21}$ POT & 2.7$\times$10$^{22}$ POT & 2.5$\times$10$^{20}$ $\mu$ decays \\ \hline
\rule{0pt}{3ex}Energy range~[GeV] & 0.5--8 & 0.1--1.2 & 1--25 \\ \hline
\rule{0pt}{3ex}Baseline length~[km] & 1300 & 295 & 1736 \\ \hline
\rule{0pt}{3ex}\multirow{2}{*}{Target material} & \multirow{2}{*}{liquid argon} & \multirow{2}{*}{ultra-pure water} & magnetized iron and \\ 
\rule{0pt}{3ex} &  &  & emulsion hybrid \\ \hline
\rule{0pt}{3ex}Detector size~[kton] & 40 & 187(374) & 50 \\ \hline
\rule{0pt}{3ex}Reference & Ref.~\cite{Abi:2020kei} & {\color{black}Ref.~\cite{Abe:2015zbg,Du:2021rdg}} & this work \\
\hline
\end{tabular}
\end{center}
\end{table}
In this work, we studied the prospects of building an accelerator-based neutrino oscillation experiment in China. We presented a survey of five accelerator laboratories and two underground laboratories capable of hosting neutrino beam and detector facilities. The physics potential of the considered laboratories were investigated according to their geographical locations by simulating a long-baseline neutrino experiment. The study presented in this work focuses on {\color{black}four} aspects of neutrino oscillations: (1) CP violation and precision measurement of Dirac {\em CP} phase $\delta_{CP}$, (2) precision measurement of the oscillation parameters $\theta_{23}$ and $\Delta m_{31}^2$, (3) unitarity of the neutrino mixing matrix and (4) sensitivity to non-standard interactions in the neutrino sector.

In the center of our work is discussion on the design of future accelerator neutrino experiment that could be realized in China. Focusing on long-baseline neutrino oscillations, we presented a comparative study of the baseline lengths that could be accessible in the considered laboratories. For the neutrino beam, we studied two alternative scenarios, where one is driven by muon decay and the other by beta decays. Our results show configurations with baseline lengths 800--1400~km provide favourable conditions for nearly all measurement goals in both beam configurations. Three of the considered laboratory pairs (Nanjing$\rightarrow$JUNO, CiADS$\rightarrow$CJPL and CSNS$\rightarrow$CJPL) fall into this category. Exceptions are found with the non-unitarity parameter $\alpha_{32}$ and non-standard interaction parameter $\epsilon^m_{\mu \tau}$ where baseline lengths 1600--1800~km are more beneficial. Three configurations (Nanjing$\rightarrow$CJPL, SPPC$\rightarrow$CJPL and CAS-IMP$\rightarrow$JUNO) belong to the latter category.

As a concrete example of what could be a future accelerator neutrino experiment in China, we present case study on a configuration where a muon-decay-based neutrino beam is used over at the SPPC$\rightarrow$CJPL baseline. The baseline length of this configuration is 1736~km. We found that 25~GeV muon beam and 50~kton neutrino detector based on the magnetized iron and emulsion cloud chamber technologies provide competitive sensitivities to all physics goals. A summary of the expected sensitivities is provided in table\,\ref{tab:nsi_prosp}, where sensitivities obtained for the baseline setup of PROMPT are presented. The expected sensitivities are also provided for DUNE and T2HK. We note that the sensitivity to tau neutrino appearance channel $\nu_e \rightarrow \nu_\tau$ in the PROMPT setup contributes significantly to the sensitivity to $\epsilon^m_{e \tau}$ but its effect is only modest in other measurements. {\color{black}The effects of tau neutrino appearance channels are presently limited by the relatively low statistics allowed by the existing technologies.}

We find altogether very promising prospects for an accelerator neutrino experiment in China. Our survey reveals a very promising landscape for future neutrino beam and detector facilities thanks to the growing research infrastructure.

\acknowledgments
SV thanks Pedro Pasquini for helpful discussions on non-unitary neutrino mixing. The authors were supported in part by National Natural Science Foundation of China under Grant No. 12075326 and No. 11881240247, by Guangdong Basic and Applied Basic Research Foundation under Grant No. 2019A1515012216. SV was also supported by China Postdoctoral Science Foundation under Grant No.\,2020M672930. JT acknowledges the support from the CAS Center for Excellence in Particle Physics (CCEPP).

\bibliographystyle{CTP}
\bibliography{References}

\end{document}